\documentclass[%
 reprint,
 amsmath,amssymb,
 aps,
]{revtex4-2}
\usepackage{graphicx, color}% Include figure files
\usepackage{dcolumn}% Align table columns on decimal point
\usepackage{bm}% bold math

\usepackage{booktabs}
\usepackage{float}
\usepackage{algorithm}
\usepackage{algorithmic}
\usepackage{xcolor}

\usepackage[colorlinks,urlcolor=blue,linkcolor=blue,anchorcolor=blue,citecolor=blue]{hyperref} %refernce and surper-link

\begin{document}

\preprint{APS/123-QED}

\title{Superconductivity in atom-intercalated quaternary hydrides under ambient pressure}% Force line breaks with \\

%Exploring the Potential for Ambient Pressure Superconductivity Enhancement via Atom Intercalation in Quaternary Hydrides

\author{Bo-Wen Yao$^{1,2}$}
\thanks{These authors contributed equally to this work.}
\author{Zhenfeng Ouyang$^{1,2}$}
\thanks{These authors contributed equally to this work.}
\author{Xiao-Qi Han$^{1,2}$}
\thanks{These authors contributed equally to this work.}
\author{Chang-Jiang Wu$^{1,2}$}
\author{Peng-Jie Guo$^{1,2}$}
\author{Ze-Feng Gao$^{1,2}$}
\email{zfgao@ruc.edu.cn}
\author{Zhong-Yi Lu$^{1,2,3}$}
\email{zlu@ruc.edu.cn}
\affiliation{1. School of Physics and Beijing Key Laboratory of Opto-electronic Functional Materials $\&$ Micro-nano Devices. Renmin University of China, Beijing 100872, China}
\affiliation{2. Key Laboratory of Quantum State Construction and Manipulation (Ministry of Education), Renmin University of China, Beijing 100872, China}
\affiliation{3. Hefei National Laboratory, Hefei 230088, China}

\begin{abstract}

Hydrogen-rich materials are the most promising candidates for high-temperature conventional superconductors under ambient pressure. Multinary hydrides have abundant structural configurations and are more promising to find high-temperature superconductors at ambient pressure, but searching for multinary materials in complex phase space is a great challenge. In this work, we used our developed AI search engine (InvDesFlow) to perform extensive investigations regarding ambient stable superconducting hydrides. Several quaternary hydrides with high superconducting temperature~($T_c$) are predicted. In particular, the superconducting $T_c$ of K$_2$GaCuH$_6$ and K$_2$LiCuH$_6$ are calculated to be 68~K and 53~K under ambient pressure, respectively, which shows a significant enhancement in comparison with that of K$_2$CuH$_6$~($T_c$ $\sim$ 16~K). We also find that intercalating atoms could cause phonon softening and induce more phonon modes with strong electron-phonon coupling. Hence, we propose that intercalating atoms is a feasible approach in searching for superconducting quaternary hydrides.
\end{abstract}

\maketitle

\section{Introduction}
Since the first superconductivity was discovered in 1911~\cite{onnes1911comm}, superconductors with high critical temperature~($T_c$) under ambient pressure have been the target of efforts to find. In the exploration of conventional superconductors, MgB$_2$ currently holds the record for the highest $T_c$ among successfully synthesized materials under ambient pressure, with a $T_c$ of 39 K~\cite{nagamatsu2001superconductivity}, which does not surpass the McMillan limit~\cite{mcmillan1968transition}. According to the Bardeen-Cooper-Schrieffer~(BCS) theory~\cite{bardeen1957theory}, lighter elements typically exhibit higher Debye frequencies. Thus, Ashcroft \emph{et al.} proposed that metallic hydrogen could exhibit a high $T_c$~\cite{ashcroft1968metallic}. However, inducing hydrogen into a metallic phase requires extremely high pressures, posing significant experimental challenges~\cite{babaev2004superconductor}. In hydrides, hydrogen atoms are precompressed within crystal lattices, which could lower the extreme environmental conditions needed for the metalization of hydrogen~\cite{ashcroft2004hydrogen}. With the advancement of high-pressure techniques, many high-$T_c$ superconducting hydrides have been reported~\cite{wang2012superconductive,ma2022high,drozdov2015conventional,li2014metallization}.

Early studies primarily focused on binary hydrides, which are relatively easy to design, synthesize, and characterize due to their simple compositions, for example, CaH$_6$~\cite{wang2012superconductive}($T_c$ = 205~K at 172~GPa), H$_3$S~\cite{drozdov2015conventional}~($T_c$ = 203~K at 220~GPa), and LaH$_{10}$~\cite{liu2017potential,drozdov2019superconductivity}($T_c$ = 250~K at 170~GPa). However, the limited structural diversity of binary hydrides has constrained further exploration, shifting research interest increasingly toward ternary hydrides~\cite{flores2020perspective,lv2020theory,zhao2024superconducting}. Recently, the ternary hydride LaBeH$_8$ was successfully synthesized under high pressures of 110–130 GPa, exhibiting a superconducting $T_c$ of 110~K at 80~GPa~\cite{song2023stoichiometric}. Similar result was also found in La-Ce-H compounds~\cite{chen2023enhancement}, ThBeH$_8$ and CeBeH$_8$~\cite{jiang2024ternary}, LaBH$_8$~\cite{di2021bh,liang2021prediction}, SrSiH$_8$, BaSiH$_8$~\cite{lucrezi2022silico} and a broader class of $AX$H$_8$ materials (where $A$ = Sc, Ca, Y, Sr, La, Ba and $X$ = Be, B, Al)~\cite{zhang2022design}.

Recent efforts have identified promising ternary superconducting hydride candidates, such as Mg$_2$XH$_6$ (X = Rh, Ir, Pd, Pt)~\cite{sanna2024prediction} and double perovskite hydrides~\cite{cerqueira2024searching}. Among them, Mg$_2$IrH$_6$ was predicted to be superconducting under ambient pressure with a $T_c$ of $\sim$ 160~K~\cite{dolui2024feasible}. Using the AI search engine, InvDesFlow, our recent work proposed a 216-type ternary hydride Li$_2$AuH$_6$ with a superconducting $T_c$ of 140~K under ambient pressure~\cite{ouyang2025strong}. The strong electron-phonon coupling~(EPC) of Li$_2$AuH$_6$ is mainly contributed by the Au-H octahedron, which is hence proposed to be a BCS superconducting unit. Further studies regarding Li$_2$AuH$_6$ suggested that intercalating extra atoms may further enhance EPC, which motivates us to try to search for superconducting quaternary hydrides. And the rising AI technology just provides a suitable approach to investigate quaternary hydride.

In this work, using our AI search engine, InvDesFlow~\cite{xiao2025invdesflow,InvDesFlow-AL}, we propose atom-intercalated quaternary hydrides based on the 216-type ternary hydrides. The detailed workflow of our AI method with process diagram is provided in Supplementary Materials~(SM)~\cite{SM} Section I, the specific setting of the AI model can be found in the source code~\cite{code}. By performing the density functional theory~(DFT)~\cite{hohenberg1964inhomogeneous, kohn1965self} and superconducting EPC studies, we find that the intercalated atoms bring modulation for phonon, EPC, and superconducting $T_c$ of quaternary hydrides. Typically, the superconducting $T_c$ of K$_2$GaCuH$_6$ and K$_2$LiCuO$_6$ are respectively predicted to be $\sim$ 68~K and 53~K under ambient pressure, which is significantly enhanced in comparison with that of K$_2$CuH$_6$~(16~K).

\section{Methods}

In our calculations, we performed first-principles electronic structure calculations with the QUANTUM-ESPRESSO package~\cite{giannozzi2009quantum}. The exchange and correlation functional was treated with the Perdew-Burke-Ernzerhof~(PBE) generalized gradient approximation~(GGA)~\cite{perdew1996generalized}, while electron-ion interactions were modeled using norm-conserving Vanderbilt pseudopotentials~\cite{hamann2013optimized}. For all hydrogen-rich materials, we adopted a plane-wave kinetic energy cutoff of 80~Ry and a charge density cutoff of 320~Ry. The charge densities were calculated on an unshifted 16$\times$16$\times$16 $\textbf{k}$-point grid. The Methfessel-Paxton smearing method~\cite{methfessel1989high} was applied with an energy width of 0.02~Ry. According to density functional perturbation theory~\cite{baroni2001phonons}, the dynamical matrices and perturbation potentials were evaluated on a $\Gamma$-centered 4$\times$4$\times$4 \textbf{q}-point grid. A 4$\times$4$\times$4 $\textbf{k}$-point grid in the Brillouin zone was used to construct the maximally localized Wannier functions~(MLWFs)~\cite{pizzi2020wannier90}. The Electron-Phonon Wannier(EPW) package~\cite{ponce2016epw} was employed to evaluate the EPC constant $\lambda$ through a convergence test within fine electron~(64$\times$64$\times$64) and phonon~(16$\times$16$\times$16) grids. A Gaussian smearing of 0.5~meV was applied to the phonon Dirac $\delta$ functions. Ultimately, the $T_c$ of each crystal was obtained by solving the anisotropic Eliashberg equations~\cite{ponce2016epw,choi2003anisotropic,margine2013anisotropic}. The Matsubara frequencies were truncated at eight times the maximum phonon energy in each material.

The mode- and wavevector-dependent coupling $\lambda_{\textbf{q}\nu}$ can be computed using the following expression
\begin{eqnarray}
    \lambda_{\textbf{q}\nu} = \frac{2}{\hbar N(0)N_\textbf{k}}\sum _{nm\textbf{k}}\frac{1}{\omega_{\textbf{q}\nu}} \left| g^{nm}_{\textbf{k},\textbf{q}\nu}\right|^2\delta(\varepsilon^n_{\textbf{k}})\delta(\varepsilon^n_{\textbf{k}+\textbf{q}}),
\end{eqnarray}
where $N(0)$ is the Fermi level density of states~(DOS) of electrons, $N_\textbf{k}$ is the sum of \textbf{k} points in the fine kmesh. The phonon frequency is expressed as $\omega_{\textbf{q}\nu}$. The EPC matrix element is denoted as $g^{nm}_{\textbf{k},\textbf{q}\nu}$. The indices of energy bands and phonon modes are written as $(n,m)$ and $\nu$. $\varepsilon^n_{\textbf{k}}$ and $\varepsilon^n_{\textbf{k}+\textbf{q}}$ denote the Kohn-Sham orbital eigenvalues measured from the Fermi level.

The EPC constant $\lambda$ can be calculated by summing  $\lambda_{\textbf{q}\nu}$ over the entire first Brillouin space or by integrating the Eliashberg spectral function $\alpha^2F(\omega)$,
\begin{eqnarray}
    \lambda = \frac{1}{N_\textbf{q}}\sum_{\textbf{q}\nu}\lambda_{\textbf{q}\nu}=2\int \frac{\alpha^2F(\omega)}{\omega}d\omega,
\end{eqnarray}
where $N_\textbf{q}$ denotes the total number of $\textbf{q}$ points within the fine $\textbf{q}$ mesh, and the Eliashberg spectral function $\alpha^2F(\omega)$ is computed with
\begin{eqnarray}
    \alpha^2F(\omega)=\frac{1}{2N_\textbf{q}}\sum_{\textbf{q}\nu}\lambda_{\textbf{q}\nu}\omega_{\textbf{q}\nu}\delta(\omega-\omega_{\textbf{q}\nu}).
\end{eqnarray}

\section{Results}

\begin{figure}[htbp]
    \centering
    \includegraphics[width=1\linewidth]{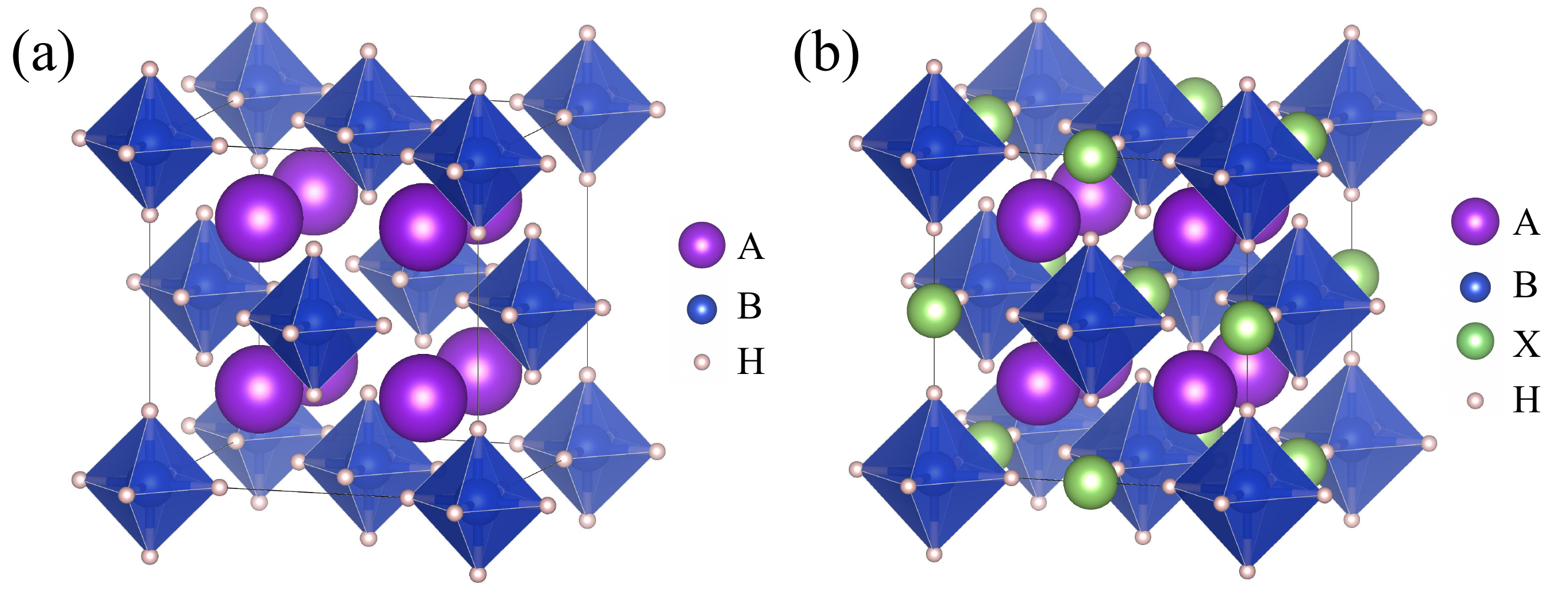}
    \caption{Crystal structure of (a)  $A_2B$H$_6$~($A$ = K and Na; $B$ = Cu and Ag), (b) $A_2XB$H$_6$~($A$ = K and Na; $B$ = Cu and Ag; $X$ = Ga and Li), respectively.}
    \label{fig:lattice}
\end{figure}

As shown in Fig.~\ref{fig:lattice}, all multinary hydrides $A_2B$H$_6$~($A$ = K and Na; $B$ = Cu and Ag) and $A_2XB$H$_6$~($A$ = K and Na; $B$ = Cu and Ag; $X$ = Ga and Li) in our research share the same space group of  $Fm\overline{3}m$. $A$, $B$, and H atoms occupy 8$c$~(0.25, 0.25, 0.25), 4$a$~(0, 0, 0), and 24$e$~(0.25, 0, 0) Wyckoff positions, respectively, where $B$ locates at the center of an octahedron that is constructed by the six nearest hydrogen atoms. In quaternary hydride, the intercalated $X$ atoms occupy the position of 4$b$~(0.5, 0, 0).

\begin{figure}[htbp]
    \centering
    \includegraphics[width=1\linewidth]{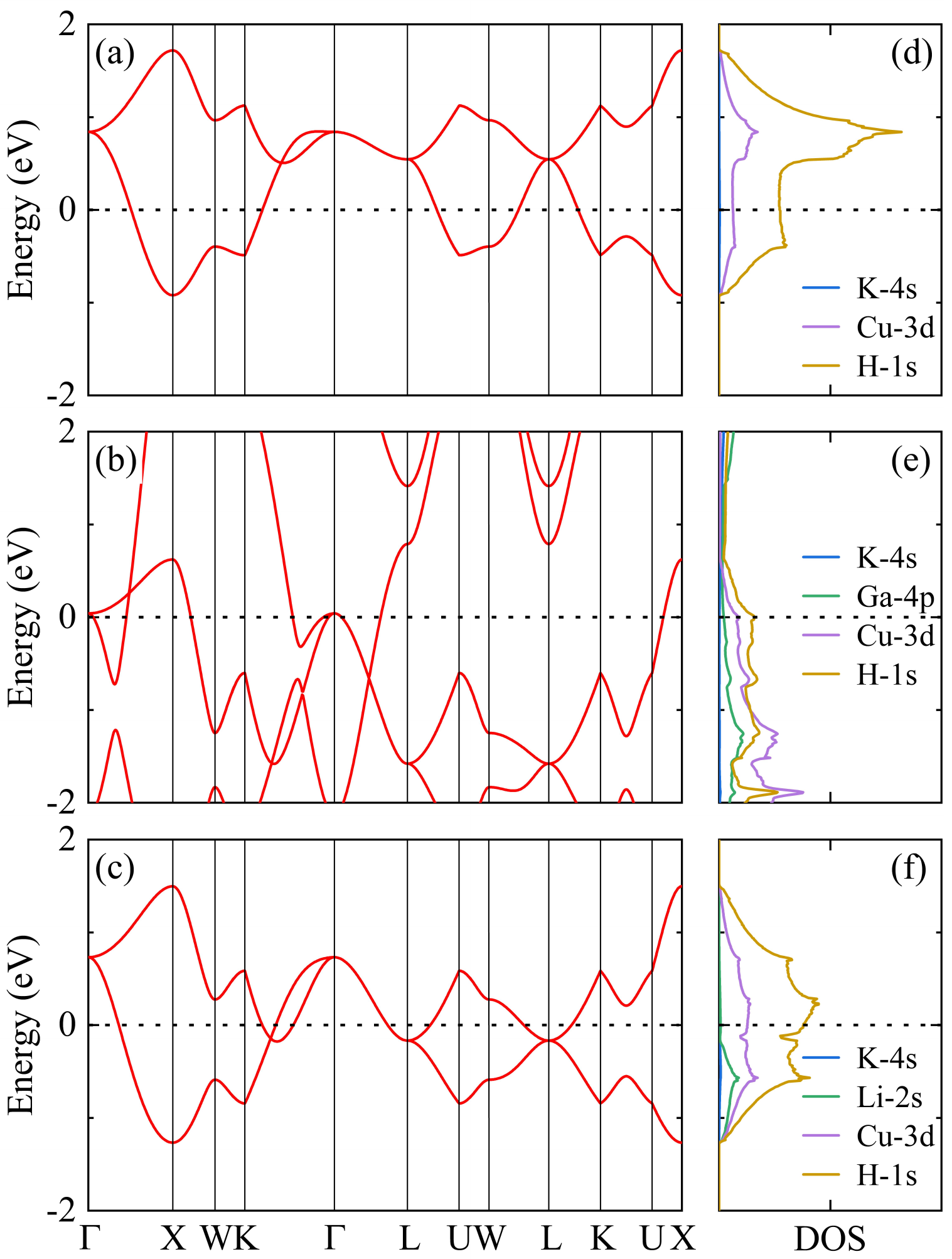}
    \caption{(a-c)~Band structures and (d-f)~atomic orbital-projected density of states of K$_2$CuH$_6$, K$_2$GaCuH$_6$, and K$_2$LiCuH$_6$, respectively. The Fermi level is set to zero.}
    \label{fig:k2-cu-h_band}
\end{figure}

\begin{figure}[htbp]
    \centering
    \includegraphics[width=1\linewidth]{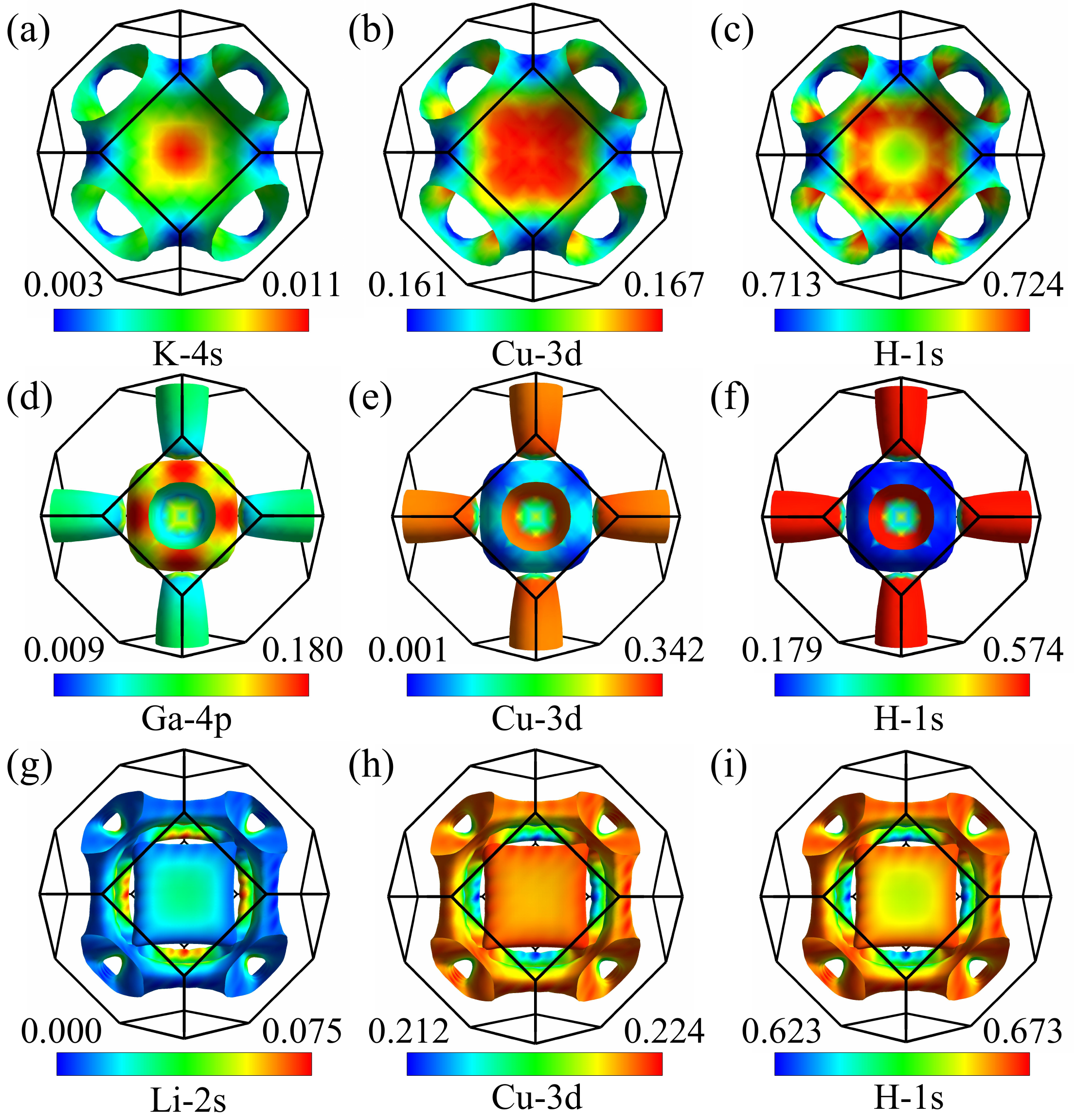}
    \caption{The Fermi surfaces with colored representations of orbital-projected weights of (a-c)~K$_2$CuH$_6$, (d-f)~K$_2$GaCuH$_6$, and (g-i)~K$_2$LiCuH$_6$, respectively. Contribution of K atom is not shown because there orbital-projected weight on the Fermi surface is weaker than that of intercalated atom.}
    \label{fig:k2-cu-h_fs}
\end{figure}

Figure~\ref{fig:k2-cu-h_band} shows the electronic structures of K$_2$CuH$_6$, K$_2$GaCuH$_6$, and K$_2$LiCuH$_6$ under ambient pressure. We see that K$_2$CuH$_6$ is a single-band metal, K$_2$GaCuH$_6$ and K$_2$LiCuH$_6$ each have two bands crossing the Fermi level. Figures~\ref{fig:k2-cu-h_band}(d-f) show the corresponding calculated projected density of states~(PDOS). The bands near the Fermi level of these hydrides are primarily contributed by Cu-H octahedrons, with negligible contributions from the K atom. A similar situation is also found in the intercalated atoms in the two quaternary hydrides, Ga and Li atoms contribute little to the low-energy PDOS.

Figures~\ref{fig:k2-cu-h_fs}(a-c) show the orbital-projected Fermi surfaces of K$_2$CuH$_6$. Since the electronegativity of the K atom is weaker than that of H and Cu, the K-4$s$ orbital contributes little, and the electronic states on the Fermi surface mainly come from the Cu-H octahedron. Meanwhile, the H-1$s$ orbital in the octahedron is dominant because to H atom is more electronegative. In quaternary hydrides~[Figs.~\ref{fig:k2-cu-h_fs}(d-f) and Figs.~\ref{fig:k2-cu-h_fs}(g-i)]. Since Ga and Li are less electronegative than Cu and H, Ga and Li cannot compete with the octahedron in their ability to capture electrons; the Fermi surfaces of the two quaternary hydrides are still mainly contributed by the H-1$s$ orbital and Cu-3$d$ orbital. It is also worth noting that the distributions of the orbital-projected weights on the Fermi surfaces show significant anisotropy. For example, H atoms contribute largely to the trumpet-shaped Fermi surface but less to the cubic Fermi surface. According to Fig.~\ref{fig:k2-cu-h_band} and Fig.~\ref{fig:k2-cu-h_fs}, it can be seen that both the number of energy bands crossing the Fermi level and the shape of the Fermi surface change significantly, indicating that the intercalated atoms have altered the electronic structures. The detailed electronic structures of the Na-Cu-H and Na-Ag-H systems are provided in SM Sections II and III, respectively.

% -----------------------------------声子结构部分----------------------------------------

\begin{figure}[htbp]
    \centering
    \includegraphics[width=1\linewidth]{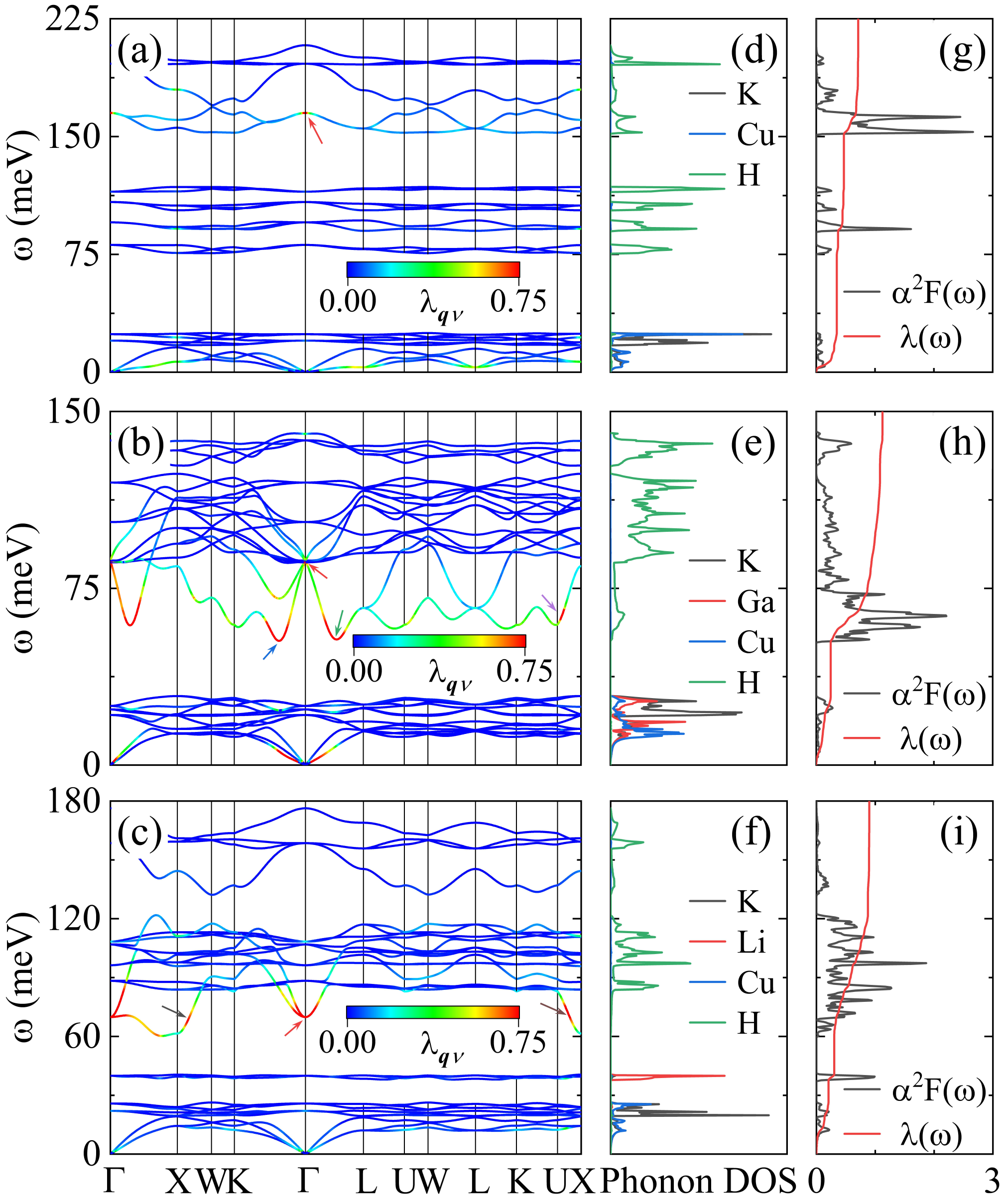}
    \caption{(a-c)~Phonon spectra with colored representations of EPC strength $\lambda_{\textbf{q}\nu}$ weights of K$_2$CuH$_6$, K$_2$GaCuH$_6$, and K$_2$LiCuH$_6$ under ambient pressure, respectively. Colored arrows denote vibration modes that contribute strong EPC. (d-f)~Projected phonon DOS and (g-i) Eliashberg spectral function $\alpha^2F(\omega)$ and accumulated $\lambda(\omega)$ of K-Cu-H system, respectively.}
    \label{fig:k2-cu-h_freq}
\end{figure}

Next, we illustrate the phonon and EPC of the K-Cu-H system under ambient pressure. Figures~\ref{fig:k2-cu-h_freq}(a-c) show the phonon spectra of K$_2$CuH$_6$, K$_2$GaCuH$_6$, and K$_2$LiCuH$_6$, respectively. There is no imaginary phonon mode under ambient pressure, which indicates that these multinary hydrides are dynamically stable. Figure~\ref{fig:k2-cu-h_freq}(d) shows the projected phonon DOS of K$_2$CuH$_6$. The phonon spectrum of K$_2$CuH$_6$ can be divided into three regions. In the region of approximately less than 30~meV, the acoustic phonons and low-frequency optical branches are mainly contributed by the vibrations of K and Cu, the rest high-energy phonons are totally contributed by the vibrations of H. As shown in Fig.~\ref{fig:k2-cu-h_freq}(e), the phonon spectrum of K$_2$GaCuH$_6$ mainly has two components. In the region below 30 meV, it is mainly contributed to by mixed vibrations of K, Ga, and Cu, and the region above 50 meV is entirely contributed to by H. The projected phonon DOS of K$_2$LiCuH$_6$ is shown in Fig.~\ref{fig:k2-cu-h_freq}(f), which has three major phonon modes contributing to it. In the region below 30 meV, phonons are mainly derived from the vibrations of K and Cu. Different from K$_2$GaCuH$_6$, around 40~meV, the Li atom contributes to the phonon DOS alone because the atomic mass of Li is significantly lighter than that of K and Cu. In the region above 60 meV, it is entirely contributed by H.

\begin{figure}[htbp]
    \centering
    \includegraphics[width=1\linewidth]{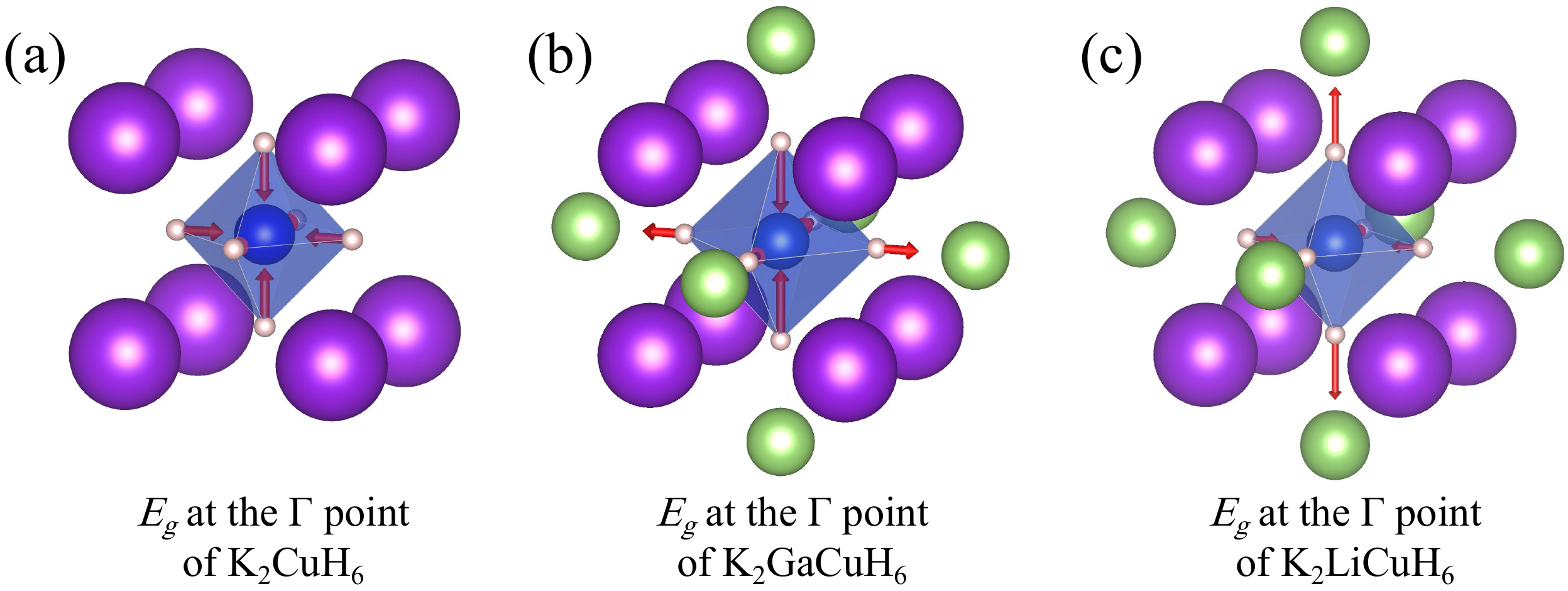}
    \caption{The vibration modes of $E_g$ at the $\Gamma$ point of (a)~K$_2$CuH$_6$, (b)~K$_2$GaCuH$_6$, and (c)~K$_2$LiCuH$_6$, respectively. The purple, blue and pink atoms denote K, Cu and H atoms, respectively; the green atom denotes the intercalated Ga or Li atom.}
    \label{fig:k2-cu-h_vibration}
\end{figure}

Moreover, the colored representations in Figs.~\ref{fig:k2-cu-h_freq}(a-c) also show the distributions of the strength of the EPC of these hydrides. In the ternary hydride K$_2$CuH$_6$, the strongest EPC is found in the $E_g$ mode~(165~meV) at the $\Gamma$ point~[indicated by the red arrow in Fig.~\ref{fig:k2-cu-h_freq}(a)]. As shown in Fig.~\ref{fig:k2-cu-h_freq}(b), after intercalating a Ga atom, the $E_g$ mode at the $\Gamma$ point still has relatively strong EPC~[indicated by the red arrow in Fig.~\ref{fig:k2-cu-h_freq}(b)], but the phonon energy decreases to about 85~meV. The intercalated Ga atom also induces more phonon modes with strong EPC, such as the mode $\sim$ 53~meV on the path from $\Gamma$ point to $K$ point that is marked by a blue arrow. Figure~\ref{fig:k2-cu-h_freq}(c) shows that the intercalated Li atom significantly enhances the EPC strength of the $E_g$ mode at the $\Gamma$ point, and the frequency drops to about 70~meV~[indicated by the red arrow in Fig.~\ref{fig:k2-cu-h_freq}(c)]. Meanwhile, two new phonon modes with strong EPC appear. In addition, it can be seen that both the intercalations of Ga and Li cause phonon softening. Figures~\ref{fig:k2-cu-h_vibration}(a-c) show the patterns of the $E_g$ mode at the $\Gamma$ point in the K-Cu-H system, and these modes with strong EPC are the breathing modes of the Cu-H octahedrons.

As shown in Figs.~\ref{fig:k2-cu-h_freq}(g-i), the EPC constants $\lambda$ of K$_2$CuH$_6$, K$_2$GaCuH$_6$, and K$_2$LiCuH$_6$ are calculated to be 0.71, 1.13, and 0.90, respectively. Figure~\ref{fig:k2-cu-h_freq}(c) shows that the phonons below 160~meV have contributed more than 80\% of the total EPC $\lambda$ in K$_2$CuH$_6$. By intercalating a Ga atom, the number of vibration modes with strong EPC increases, and these strong EPC modes ranging from 50~meV to 90~meV contribute approximately 64\% of the total $\lambda$~[Fig.~\ref{fig:k2-cu-h_freq}(h)]. A similar situation is also found in the Li-intercalated case~[Fig.~\ref{fig:k2-cu-h_freq}(i)]. These results indicate that the intercalating atoms can induce more phonon modes with strong EPC or significantly enhance the EPC strength. In addition, we show the phonon and EPC calculation results of Na-Cu-H and Na-Ag-H systems in SM Sections II and III, respectively.

\begin{figure}[htbp]
    \centering
    \includegraphics[width=1\linewidth]{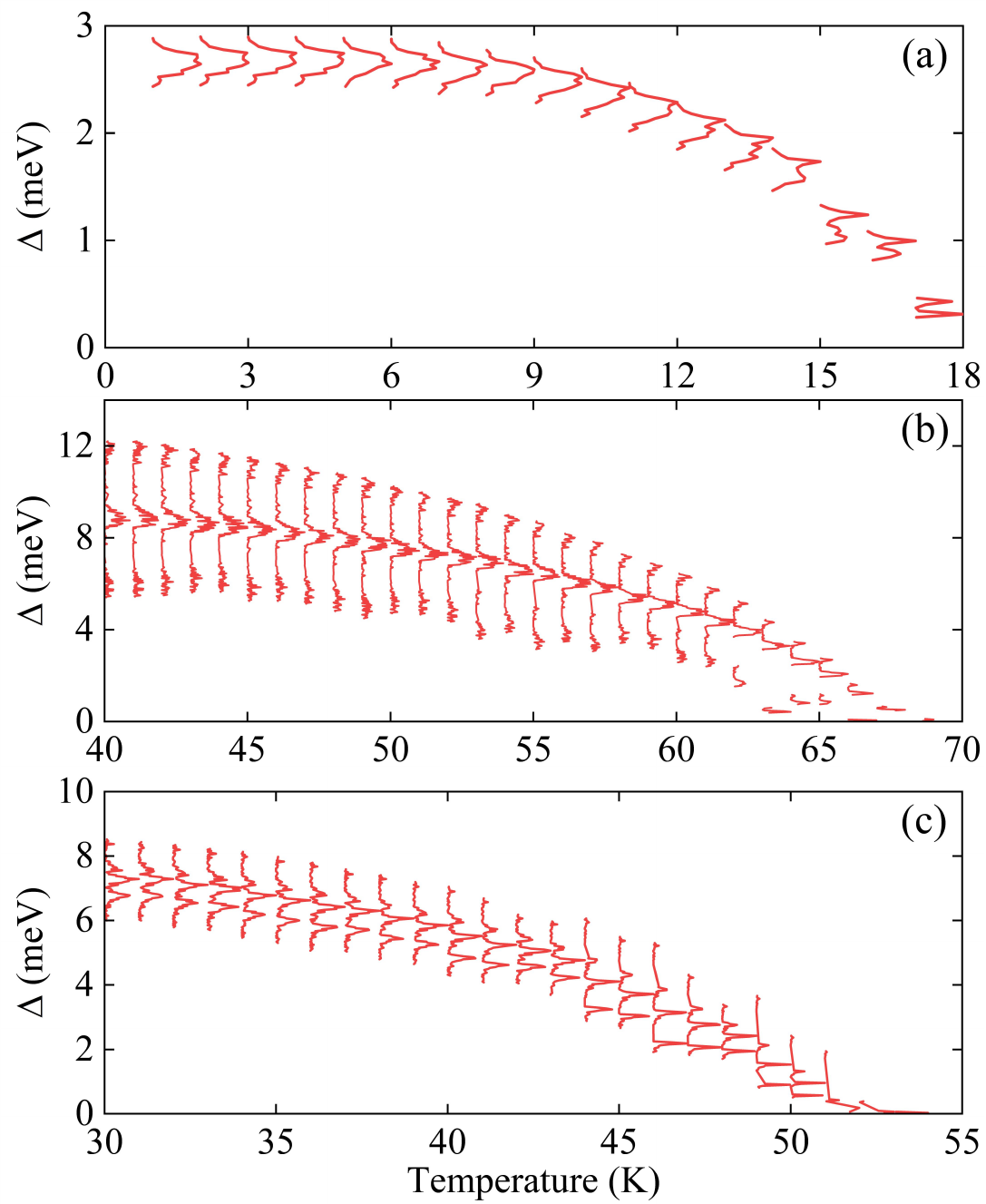}
    \caption{Normalized anisotropic superconducting gap $\Delta$ of (a)~K$_2$CuH$_6$, (b)~K$_2$GaCuH$_6$, and (c)~K$_2$LiCuH$_6$ under ambient pressure, respectively. The screening Coulomb potential $\mu ^*$ is set to be 0.1.}
    \label{fig:k-cu-h_anisotropy}
\end{figure}

By solving anisotropic Eliashberg equations, we obtain the superconducting $T_c$ of these hydrides. The superconducting gap $\Delta$ of the K-Cu-H system is shown in Fig~\ref{fig:k-cu-h_anisotropy}. The superconducting gap $\Delta$ of K$_2$CuH$_6$ disappears at 18~K, indicating that the superconducting $T_c$ of K$_2$CuH$_6$ is 18~K under ambient pressure. As for K$_2$GaCuH$_6$ and K$_2$LiCuH$_6$, the superconducting $T_c$ is respectively calculated to be 68~K and 53~K under ambient pressure~[Fig.~\ref{fig:k-cu-h_anisotropy}(b-c)], which is significantly enhanced in comparison with that of the non-intercalated K$_2$CuH$_6$.

\begin{figure}[htbp]
    \centering
    \includegraphics[width=1\linewidth]{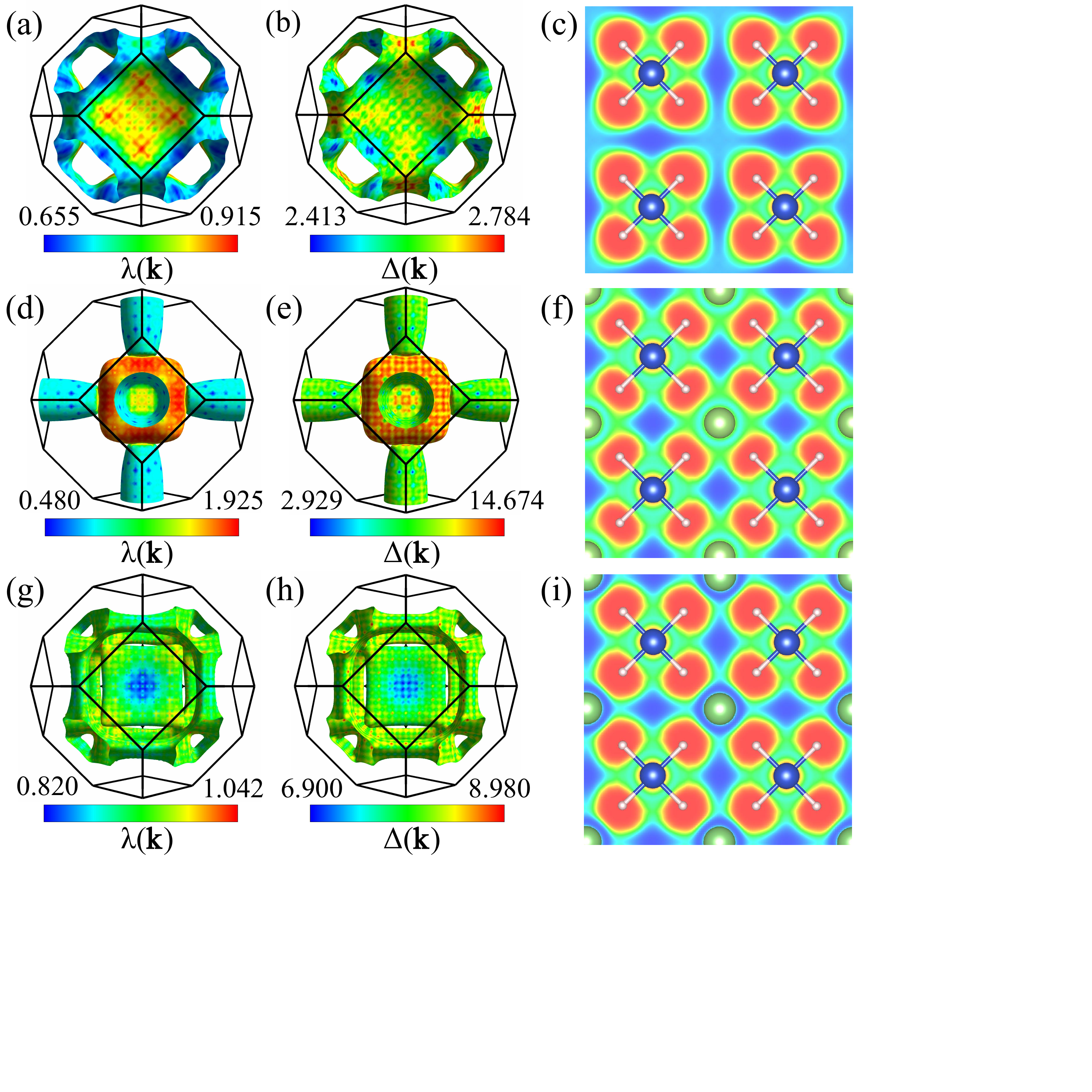}
    \caption{Distributions of EPC $\lambda$(\textbf{k}) on the Fermi surfaces, superconducting gap $\Delta$(\textbf{k}) on the Fermi surfaces, and electron localization function of (a-c)~K$_2$CuH$_6$, (d-f)~K$_2$GaCuH$_6$, and (g-i)~K$_2$LiCuH$_6$, respectively. The distributions of the superconducting gap $\Delta$(\textbf{k}) are calculated at 5~K. The blue and pink atoms denote Cu and H atoms, respectively; the green atom denotes the intercalated Ga or Li atom.}
    \label{fig:k-cu-h_elf}
\end{figure}

Figure~\ref{fig:k-cu-h_elf} shows the calculated results of anisotropic superconductivity of the K-Cu-H system. In K$_2$GaCuH$_6$, the EPC $\lambda$(\textbf{k}) on the trumpet-shaped Fermi surface is relatively weak, while the EPC $\lambda$(\textbf{k}) on the cubic edge is relatively strong~[Fig.~\ref{fig:k2-cu-h_fs}(d)]. The electron localization function~(ELF) of the crystal plane shown here passes through the plane where the Cu and H atoms are located. Electron localization functions suggest that the valence electrons are mainly localized around the H atoms. These findings suggest that the vibrations of H atoms could significantly change electronic density and induce strong EPC. The relevant calculation results of anisotropic superconductivity of Na-Cu-H and Na-Ag-H systems are shown in the SM Sections II and III, respectively.

\begin{table}[htbp]
    \caption{The EPC constant $\lambda$ and superconducting transition temperature $T_c$ by solving self-consistently the anisotropic Eliashberg equations of all multinary hydrides. The formulas with a $\dagger$ symbol in the upper right corner denote that the materials were generated by our machine learning model.}
    \begin{ruledtabular}
    \begin{tabular}{ccc}
       Formula  & EPC constant $\lambda$ & $T_c$~(K) \\ \hline
       K$_2$CuH$_6$ & 0.7120 & 16 \\
       K$_2$GaCuH$_6^\dagger$ & 1.1261 & 68 \\
       K$_2$LiCuH$_6$ & 0.8995 & 53 \\
       Na$_2$CuH$_6$ & 0.8699 & 56 \\
       Na$_2$GaCuH$_6^\dagger$ & 0.8164 & 42 \\
       Na$_2$LiCuH$_6$ & 0.7305 & 43  \\
       Na$_2$AgH$_6$ & 1.3213 & 89 \\
       Na$_2$LiAgH$_6^\dagger$ & 1.6131 & 86
    \end{tabular}
    \end{ruledtabular}
    \label{table:Tc}
\end{table}

Table~\ref{table:Tc} summarizes the calculated results of the EPC constant $\lambda$ and the superconducting $T_c$ of eight hydrides. In the K-Cu-H system, the intercalated atoms successfully enhance the strength of EPC in the quaternary hydride and significantly increase $T_c$. This indicates that the superconductivity of the crystal can be modulated by intercalating atoms. Especially when the intercalated atoms successfully induce more or stronger phonon modes with strong EPC, it is expected to further increase the superconductivity $T_c$ of the material under ambient pressure, providing a feasible direction for the discovery of more potential high-temperature superconductors.

\section{Discussion}

Due to the rich element combinations and high phase space dimensions for multinary hydrides, the design of multinary hydrides is a very difficult project, which is often time-consuming and resource-intensive in the traditional material design methods. Artificial intelligence models based on deep learning methods can effectively address this challenge. Our InvDesFlow model integrates the generative model and the discriminant model. The crystal structure can be reparameterized through the graph neural network~\cite{scarselli2008graph}, and then the feature vector is embedded into the high-dimensional latent space. The diffusion process of the generative model will train the model to learn the reasonable element composition, atomic position, and crystal structure information of a crystal, and recover the correct crystal structure from the latent space after the end of training. Different discriminant models learn crystal structure distributions from the high-dimensional latent space according to their respective tasks, and can quickly output crystal formation energy, superconducting temperature, and other information in the process of running. Since our model is an integrated pipeline, we only need to select materials with formation energy lower than the threshold and a high reference $T_c$ based on the model output, and then concentrate computing resources on those materials with greater potential, which greatly improves the efficiency of exploring potential materials.

\section{Conclusion}
In summary, combining machine learning and the first-principles theoretical calculations, we propose the atom-intercalated quaternary hydrides as high-$T_c$ superconductor candidates under ambient pressure. Specifically, the superconducting $T_c$ of intercalated K$_2$GaCuH$_6$ and K$_2$LiCuH$_6$ are respectively predicted to be $\sim$ 68~K and 53~K under ambient pressure,  which represents a significant increase compared to the 16~K of K$_2$CuH$_6$. The study of EPC shows that intercalating an atom could lead to phonon softening and induce more phonon modes with strong EPC, which may be helpful for realizing high-$T_c$ superconductivity. Our work provides a theoretical reference for searching superconducting quaternary hydrides.

\textit{Acknowledgments.} This work was financially supported by the National Natural Science Foundation of China (Grant No.62476278, No.12434009, and No.12204533). Z.Y.L. was also supported by the National Key R\&D Program of China (Grants No. 2024YFA1408601). Computational resources have been provided by the Physical Laboratory of High Performance Computing at Renmin University of China.

\bibliography{references}

\begin{thebibliography}{52}%
\makeatletter
\providecommand \@ifxundefined [1]{%
 \@ifx{#1\undefined}
}%
\providecommand \@ifnum [1]{%
 \ifnum #1\expandafter \@firstoftwo
 \else \expandafter \@secondoftwo
 \fi
}%
\providecommand \@ifx [1]{%
 \ifx #1\expandafter \@firstoftwo
 \else \expandafter \@secondoftwo
 \fi
}%
\providecommand \natexlab [1]{#1}%
\providecommand \enquote  [1]{``#1''}%
\providecommand \bibnamefont  [1]{#1}%
\providecommand \bibfnamefont [1]{#1}%
\providecommand \citenamefont [1]{#1}%
\providecommand \href@noop [0]{\@secondoftwo}%
\providecommand \href [0]{\begingroup \@sanitize@url \@href}%
\providecommand \@href[1]{\@@startlink{#1}\@@href}%
\providecommand \@@href[1]{\endgroup#1\@@endlink}%
\providecommand \@sanitize@url [0]{\catcode `\\12\catcode `\$12\catcode `\&12\catcode `\#12\catcode `\^12\catcode `\_12\catcode `\%12\relax}%
\providecommand \@@startlink[1]{}%
\providecommand \@@endlink[0]{}%
\providecommand \url  [0]{\begingroup\@sanitize@url \@url }%
\providecommand \@url [1]{\endgroup\@href {#1}{\urlprefix }}%
\providecommand \urlprefix  [0]{URL }%
\providecommand \Eprint [0]{\href }%
\providecommand \doibase [0]{https://doi.org/}%
\providecommand \selectlanguage [0]{\@gobble}%
\providecommand \bibinfo  [0]{\@secondoftwo}%
\providecommand \bibfield  [0]{\@secondoftwo}%
\providecommand \translation [1]{[#1]}%
\providecommand \BibitemOpen [0]{}%
\providecommand \bibitemStop [0]{}%
\providecommand \bibitemNoStop [0]{.\EOS\space}%
\providecommand \EOS [0]{\spacefactor3000\relax}%
\providecommand \BibitemShut  [1]{\csname bibitem#1\endcsname}%
\let\auto@bib@innerbib\@empty
%</preamble>
\bibitem [{\citenamefont {Kamerlingh~Onnes}(1911)}]{onnes1911comm}%
  \BibitemOpen
  \bibfield  {author} {\bibinfo {author} {\bibfnamefont {H.}~\bibnamefont {Kamerlingh~Onnes}},\ }\bibfield  {title} {\bibinfo {title} {The resistance of pure mercury at helium temperatures},\ }\href@noop {} {\bibfield  {journal} {\bibinfo  {journal} {Comm. Phys. Lab. Univ. Leiden}\ }\textbf {\bibinfo {volume} {122}},\ \bibinfo {pages} {124} (\bibinfo {year} {1911})}\BibitemShut {NoStop}%
\bibitem [{\citenamefont {Nagamatsu}\ \emph {et~al.}(2001)\citenamefont {Nagamatsu}, \citenamefont {Nakagawa}, \citenamefont {Muranaka}, \citenamefont {Zenitani},\ and\ \citenamefont {Akimitsu}}]{nagamatsu2001superconductivity}%
  \BibitemOpen
  \bibfield  {author} {\bibinfo {author} {\bibfnamefont {J.}~\bibnamefont {Nagamatsu}}, \bibinfo {author} {\bibfnamefont {N.}~\bibnamefont {Nakagawa}}, \bibinfo {author} {\bibfnamefont {T.}~\bibnamefont {Muranaka}}, \bibinfo {author} {\bibfnamefont {Y.}~\bibnamefont {Zenitani}},\ and\ \bibinfo {author} {\bibfnamefont {J.}~\bibnamefont {Akimitsu}},\ }\bibfield  {title} {\bibinfo {title} {Superconductivity at 39 {K} in magnesium diboride},\ }\href {https://doi.org/10.1038/35065039} {\bibfield  {journal} {\bibinfo  {journal} {Nature}\ }\textbf {\bibinfo {volume} {410}},\ \bibinfo {pages} {63} (\bibinfo {year} {2001})}\BibitemShut {NoStop}%
\bibitem [{\citenamefont {McMillan}(1968)}]{mcmillan1968transition}%
  \BibitemOpen
  \bibfield  {author} {\bibinfo {author} {\bibfnamefont {W.~L.}\ \bibnamefont {McMillan}},\ }\bibfield  {title} {\bibinfo {title} {Transition temperature of strong-coupled superconductors},\ }\href {https://doi.org/10.1103/PhysRev.167.331} {\bibfield  {journal} {\bibinfo  {journal} {Phys. Rev.}\ }\textbf {\bibinfo {volume} {167}},\ \bibinfo {pages} {331} (\bibinfo {year} {1968})}\BibitemShut {NoStop}%
\bibitem [{\citenamefont {Bardeen}\ \emph {et~al.}(1957)\citenamefont {Bardeen}, \citenamefont {Cooper},\ and\ \citenamefont {Schrieffer}}]{bardeen1957theory}%
  \BibitemOpen
  \bibfield  {author} {\bibinfo {author} {\bibfnamefont {J.}~\bibnamefont {Bardeen}}, \bibinfo {author} {\bibfnamefont {L.~N.}\ \bibnamefont {Cooper}},\ and\ \bibinfo {author} {\bibfnamefont {J.~R.}\ \bibnamefont {Schrieffer}},\ }\bibfield  {title} {\bibinfo {title} {Theory of superconductivity},\ }\href {https://doi.org/10.1103/PhysRev.108.1175} {\bibfield  {journal} {\bibinfo  {journal} {Phys. Rev.}\ }\textbf {\bibinfo {volume} {108}},\ \bibinfo {pages} {1175} (\bibinfo {year} {1957})}\BibitemShut {NoStop}%
\bibitem [{\citenamefont {Ashcroft}(1968)}]{ashcroft1968metallic}%
  \BibitemOpen
  \bibfield  {author} {\bibinfo {author} {\bibfnamefont {N.~W.}\ \bibnamefont {Ashcroft}},\ }\bibfield  {title} {\bibinfo {title} {Metallic hydrogen: A high-temperature superconductor?},\ }\href {https://doi.org/10.1103/PhysRevLett.21.1748} {\bibfield  {journal} {\bibinfo  {journal} {Phys. Rev. Lett.}\ }\textbf {\bibinfo {volume} {21}},\ \bibinfo {pages} {1748} (\bibinfo {year} {1968})}\BibitemShut {NoStop}%
\bibitem [{\citenamefont {Babaev}\ \emph {et~al.}(2004)\citenamefont {Babaev}, \citenamefont {Sudb{\o}},\ and\ \citenamefont {Ashcroft}}]{babaev2004superconductor}%
  \BibitemOpen
  \bibfield  {author} {\bibinfo {author} {\bibfnamefont {E.}~\bibnamefont {Babaev}}, \bibinfo {author} {\bibfnamefont {A.}~\bibnamefont {Sudb{\o}}},\ and\ \bibinfo {author} {\bibfnamefont {N.~W.}\ \bibnamefont {Ashcroft}},\ }\bibfield  {title} {\bibinfo {title} {A superconductor to superfluid phase transition in liquid metallic hydrogen},\ }\href {https://doi.org/10.1038/nature02910} {\bibfield  {journal} {\bibinfo  {journal} {Nature}\ }\textbf {\bibinfo {volume} {431}},\ \bibinfo {pages} {666} (\bibinfo {year} {2004})}\BibitemShut {NoStop}%
\bibitem [{\citenamefont {Ashcroft}(2004)}]{ashcroft2004hydrogen}%
  \BibitemOpen
  \bibfield  {author} {\bibinfo {author} {\bibfnamefont {N.~W.}\ \bibnamefont {Ashcroft}},\ }\bibfield  {title} {\bibinfo {title} {Hydrogen dominant metallic alloys: High temperature superconductors?},\ }\href {https://doi.org/10.1103/PhysRevLett.92.187002} {\bibfield  {journal} {\bibinfo  {journal} {Phys. Rev. Lett.}\ }\textbf {\bibinfo {volume} {92}},\ \bibinfo {pages} {187002} (\bibinfo {year} {2004})}\BibitemShut {NoStop}%
\bibitem [{\citenamefont {Wang}\ \emph {et~al.}(2012)\citenamefont {Wang}, \citenamefont {Tse}, \citenamefont {Tanaka}, \citenamefont {Iitaka},\ and\ \citenamefont {Ma}}]{wang2012superconductive}%
  \BibitemOpen
  \bibfield  {author} {\bibinfo {author} {\bibfnamefont {H.}~\bibnamefont {Wang}}, \bibinfo {author} {\bibfnamefont {J.~S.}\ \bibnamefont {Tse}}, \bibinfo {author} {\bibfnamefont {K.}~\bibnamefont {Tanaka}}, \bibinfo {author} {\bibfnamefont {T.}~\bibnamefont {Iitaka}},\ and\ \bibinfo {author} {\bibfnamefont {Y.}~\bibnamefont {Ma}},\ }\bibfield  {title} {\bibinfo {title} {Superconductive sodalite-like clathrate calcium hydride at high pressures},\ }\href {https://doi.org/10.1073/pnas.1118168109} {\bibfield  {journal} {\bibinfo  {journal} {Proc. Natl. Acad. Sci. U.S.A.}\ }\textbf {\bibinfo {volume} {109}},\ \bibinfo {pages} {6463} (\bibinfo {year} {2012})}\BibitemShut {NoStop}%
\bibitem [{\citenamefont {Ma}\ \emph {et~al.}(2022)\citenamefont {Ma}, \citenamefont {Wang}, \citenamefont {Xie}, \citenamefont {Yang}, \citenamefont {Wang}, \citenamefont {Zhou}, \citenamefont {Liu}, \citenamefont {Yu}, \citenamefont {Zhao}, \citenamefont {Wang} \emph {et~al.}}]{ma2022high}%
  \BibitemOpen
  \bibfield  {author} {\bibinfo {author} {\bibfnamefont {L.}~\bibnamefont {Ma}}, \bibinfo {author} {\bibfnamefont {K.}~\bibnamefont {Wang}}, \bibinfo {author} {\bibfnamefont {Y.}~\bibnamefont {Xie}}, \bibinfo {author} {\bibfnamefont {X.}~\bibnamefont {Yang}}, \bibinfo {author} {\bibfnamefont {Y.}~\bibnamefont {Wang}}, \bibinfo {author} {\bibfnamefont {M.}~\bibnamefont {Zhou}}, \bibinfo {author} {\bibfnamefont {H.}~\bibnamefont {Liu}}, \bibinfo {author} {\bibfnamefont {X.}~\bibnamefont {Yu}}, \bibinfo {author} {\bibfnamefont {Y.}~\bibnamefont {Zhao}}, \bibinfo {author} {\bibfnamefont {H.}~\bibnamefont {Wang}}, \emph {et~al.},\ }\bibfield  {title} {\bibinfo {title} {High-temperature superconducting phase in clathrate calcium hydride {CaH}$_6$ up to 215 {K} at a pressure of 172 {GPa}},\ }\href {https://doi.org/10.1103/PhysRevLett.128.167001} {\bibfield  {journal} {\bibinfo  {journal} {Phys. Rev. Lett.}\ }\textbf {\bibinfo {volume} {128}},\ \bibinfo {pages} {167001} (\bibinfo {year} {2022})}\BibitemShut {NoStop}%
\bibitem [{\citenamefont {Drozdov}\ \emph {et~al.}(2015)\citenamefont {Drozdov}, \citenamefont {Eremets}, \citenamefont {Troyan}, \citenamefont {Ksenofontov},\ and\ \citenamefont {Shylin}}]{drozdov2015conventional}%
  \BibitemOpen
  \bibfield  {author} {\bibinfo {author} {\bibfnamefont {A.~P.}\ \bibnamefont {Drozdov}}, \bibinfo {author} {\bibfnamefont {M.~I.}\ \bibnamefont {Eremets}}, \bibinfo {author} {\bibfnamefont {I.~A.}\ \bibnamefont {Troyan}}, \bibinfo {author} {\bibfnamefont {V.}~\bibnamefont {Ksenofontov}},\ and\ \bibinfo {author} {\bibfnamefont {S.~I.}\ \bibnamefont {Shylin}},\ }\bibfield  {title} {\bibinfo {title} {Conventional superconductivity at 203 kelvin at high pressures in the sulfur hydride system},\ }\href {https://doi.org/10.1038/nature14964} {\bibfield  {journal} {\bibinfo  {journal} {Nature}\ }\textbf {\bibinfo {volume} {525}},\ \bibinfo {pages} {73} (\bibinfo {year} {2015})}\BibitemShut {NoStop}%
\bibitem [{\citenamefont {Li}\ \emph {et~al.}(2014)\citenamefont {Li}, \citenamefont {Hao}, \citenamefont {Liu}, \citenamefont {Li},\ and\ \citenamefont {Ma}}]{li2014metallization}%
  \BibitemOpen
  \bibfield  {author} {\bibinfo {author} {\bibfnamefont {Y.}~\bibnamefont {Li}}, \bibinfo {author} {\bibfnamefont {J.}~\bibnamefont {Hao}}, \bibinfo {author} {\bibfnamefont {H.}~\bibnamefont {Liu}}, \bibinfo {author} {\bibfnamefont {Y.}~\bibnamefont {Li}},\ and\ \bibinfo {author} {\bibfnamefont {Y.}~\bibnamefont {Ma}},\ }\bibfield  {title} {\bibinfo {title} {The metallization and superconductivity of dense hydrogen sulfide},\ }\href {https://doi.org/10.1063/1.4874158} {\bibfield  {journal} {\bibinfo  {journal} {J. Chem. Phys.}\ }\textbf {\bibinfo {volume} {140}},\ \bibinfo {pages} {174712} (\bibinfo {year} {2014})}\BibitemShut {NoStop}%
\bibitem [{\citenamefont {Liu}\ \emph {et~al.}(2017)\citenamefont {Liu}, \citenamefont {Naumov}, \citenamefont {Hoffmann}, \citenamefont {Ashcroft},\ and\ \citenamefont {Hemley}}]{liu2017potential}%
  \BibitemOpen
  \bibfield  {author} {\bibinfo {author} {\bibfnamefont {H.}~\bibnamefont {Liu}}, \bibinfo {author} {\bibfnamefont {I.~I.}\ \bibnamefont {Naumov}}, \bibinfo {author} {\bibfnamefont {R.}~\bibnamefont {Hoffmann}}, \bibinfo {author} {\bibfnamefont {N.~W.}\ \bibnamefont {Ashcroft}},\ and\ \bibinfo {author} {\bibfnamefont {R.~J.}\ \bibnamefont {Hemley}},\ }\bibfield  {title} {\bibinfo {title} {Potential high-{$T_c$} superconducting lanthanum and yttrium hydrides at high pressure},\ }\href {https://doi.org/10.1073/pnas.1704505114} {\bibfield  {journal} {\bibinfo  {journal} {Proc. Natl. Acad. Sci. U.S.A.}\ }\textbf {\bibinfo {volume} {114}},\ \bibinfo {pages} {6990} (\bibinfo {year} {2017})}\BibitemShut {NoStop}%
\bibitem [{\citenamefont {Drozdov}\ \emph {et~al.}(2019)\citenamefont {Drozdov}, \citenamefont {Kong}, \citenamefont {Minkov}, \citenamefont {Besedin}, \citenamefont {Kuzovnikov}, \citenamefont {Mozaffari}, \citenamefont {Balicas}, \citenamefont {Balakirev}, \citenamefont {Graf}, \citenamefont {Prakapenka} \emph {et~al.}}]{drozdov2019superconductivity}%
  \BibitemOpen
  \bibfield  {author} {\bibinfo {author} {\bibfnamefont {A.~P.}\ \bibnamefont {Drozdov}}, \bibinfo {author} {\bibfnamefont {P.~P.}\ \bibnamefont {Kong}}, \bibinfo {author} {\bibfnamefont {V.~S.}\ \bibnamefont {Minkov}}, \bibinfo {author} {\bibfnamefont {S.~P.}\ \bibnamefont {Besedin}}, \bibinfo {author} {\bibfnamefont {M.~A.}\ \bibnamefont {Kuzovnikov}}, \bibinfo {author} {\bibfnamefont {S.}~\bibnamefont {Mozaffari}}, \bibinfo {author} {\bibfnamefont {L.}~\bibnamefont {Balicas}}, \bibinfo {author} {\bibfnamefont {F.~F.}\ \bibnamefont {Balakirev}}, \bibinfo {author} {\bibfnamefont {D.~E.}\ \bibnamefont {Graf}}, \bibinfo {author} {\bibfnamefont {V.~B.}\ \bibnamefont {Prakapenka}}, \emph {et~al.},\ }\bibfield  {title} {\bibinfo {title} {Superconductivity at 250 {K} in lanthanum hydride under high pressures},\ }\href {https://doi.org/10.1038/s41586-019-1201-8} {\bibfield  {journal} {\bibinfo  {journal} {Nature}\ }\textbf {\bibinfo {volume} {569}},\ \bibinfo {pages} {528} (\bibinfo {year} {2019})}\BibitemShut
  {NoStop}%
\bibitem [{\citenamefont {Flores-Livas}\ \emph {et~al.}(2020)\citenamefont {Flores-Livas}, \citenamefont {Boeri}, \citenamefont {Sanna}, \citenamefont {Profeta}, \citenamefont {Arita},\ and\ \citenamefont {Eremets}}]{flores2020perspective}%
  \BibitemOpen
  \bibfield  {author} {\bibinfo {author} {\bibfnamefont {J.~A.}\ \bibnamefont {Flores-Livas}}, \bibinfo {author} {\bibfnamefont {L.}~\bibnamefont {Boeri}}, \bibinfo {author} {\bibfnamefont {A.}~\bibnamefont {Sanna}}, \bibinfo {author} {\bibfnamefont {G.}~\bibnamefont {Profeta}}, \bibinfo {author} {\bibfnamefont {R.}~\bibnamefont {Arita}},\ and\ \bibinfo {author} {\bibfnamefont {M.}~\bibnamefont {Eremets}},\ }\bibfield  {title} {\bibinfo {title} {A perspective on conventional high-temperature superconductors at high pressure: Methods and materials},\ }\href {https://doi.org/10.1016/j.physrep.2020.02.003} {\bibfield  {journal} {\bibinfo  {journal} {Phys. Rep.}\ }\textbf {\bibinfo {volume} {856}},\ \bibinfo {pages} {1} (\bibinfo {year} {2020})}\BibitemShut {NoStop}%
\bibitem [{\citenamefont {Lv}\ \emph {et~al.}(2020)\citenamefont {Lv}, \citenamefont {Sun}, \citenamefont {Liu},\ and\ \citenamefont {Ma}}]{lv2020theory}%
  \BibitemOpen
  \bibfield  {author} {\bibinfo {author} {\bibfnamefont {J.}~\bibnamefont {Lv}}, \bibinfo {author} {\bibfnamefont {Y.}~\bibnamefont {Sun}}, \bibinfo {author} {\bibfnamefont {H.}~\bibnamefont {Liu}},\ and\ \bibinfo {author} {\bibfnamefont {Y.}~\bibnamefont {Ma}},\ }\bibfield  {title} {\bibinfo {title} {Theory-oriented discovery of high-temperature superconductors in superhydrides stabilized under high pressure},\ }\href {https://doi.org/10.1063/5.0033232} {\bibfield  {journal} {\bibinfo  {journal} {Matter Radiat. Extrem.}\ }\textbf {\bibinfo {volume} {5}},\ \bibinfo {pages} {068101} (\bibinfo {year} {2020})}\BibitemShut {NoStop}%
\bibitem [{\citenamefont {Zhao}\ \emph {et~al.}(2024)\citenamefont {Zhao}, \citenamefont {Huang}, \citenamefont {Zhang}, \citenamefont {Chen}, \citenamefont {Du}, \citenamefont {Duan},\ and\ \citenamefont {Cui}}]{zhao2024superconducting}%
  \BibitemOpen
  \bibfield  {author} {\bibinfo {author} {\bibfnamefont {W.}~\bibnamefont {Zhao}}, \bibinfo {author} {\bibfnamefont {X.}~\bibnamefont {Huang}}, \bibinfo {author} {\bibfnamefont {Z.}~\bibnamefont {Zhang}}, \bibinfo {author} {\bibfnamefont {S.}~\bibnamefont {Chen}}, \bibinfo {author} {\bibfnamefont {M.}~\bibnamefont {Du}}, \bibinfo {author} {\bibfnamefont {D.}~\bibnamefont {Duan}},\ and\ \bibinfo {author} {\bibfnamefont {T.}~\bibnamefont {Cui}},\ }\bibfield  {title} {\bibinfo {title} {Superconducting ternary hydrides: Progress and challenges},\ }\href {https://doi.org/10.1093/nsr/nwad307} {\bibfield  {journal} {\bibinfo  {journal} {Natl. Sci. Rev.}\ }\textbf {\bibinfo {volume} {11}},\ \bibinfo {pages} {nwad307} (\bibinfo {year} {2024})}\BibitemShut {NoStop}%
\bibitem [{\citenamefont {Song}\ \emph {et~al.}(2023)\citenamefont {Song}, \citenamefont {Bi}, \citenamefont {Nakamoto}, \citenamefont {Shimizu}, \citenamefont {Liu}, \citenamefont {Zou}, \citenamefont {Liu}, \citenamefont {Wang},\ and\ \citenamefont {Ma}}]{song2023stoichiometric}%
  \BibitemOpen
  \bibfield  {author} {\bibinfo {author} {\bibfnamefont {Y.}~\bibnamefont {Song}}, \bibinfo {author} {\bibfnamefont {J.}~\bibnamefont {Bi}}, \bibinfo {author} {\bibfnamefont {Y.}~\bibnamefont {Nakamoto}}, \bibinfo {author} {\bibfnamefont {K.}~\bibnamefont {Shimizu}}, \bibinfo {author} {\bibfnamefont {H.}~\bibnamefont {Liu}}, \bibinfo {author} {\bibfnamefont {B.}~\bibnamefont {Zou}}, \bibinfo {author} {\bibfnamefont {G.}~\bibnamefont {Liu}}, \bibinfo {author} {\bibfnamefont {H.}~\bibnamefont {Wang}},\ and\ \bibinfo {author} {\bibfnamefont {Y.}~\bibnamefont {Ma}},\ }\bibfield  {title} {\bibinfo {title} {Stoichiometric ternary superhydride {LaBeH$_8$} as a new template for high-temperature superconductivity at 110 {K} under 80 {GPa}},\ }\href {https://doi.org/10.1103/PhysRevLett.130.266001} {\bibfield  {journal} {\bibinfo  {journal} {Phys. Rev. Lett.}\ }\textbf {\bibinfo {volume} {130}},\ \bibinfo {pages} {266001} (\bibinfo {year} {2023})}\BibitemShut {NoStop}%
\bibitem [{\citenamefont {Chen}\ \emph {et~al.}(2023)\citenamefont {Chen}, \citenamefont {Huang}, \citenamefont {Semenok}, \citenamefont {Chen}, \citenamefont {Zhou}, \citenamefont {Zhang}, \citenamefont {Oganov},\ and\ \citenamefont {Cui}}]{chen2023enhancement}%
  \BibitemOpen
  \bibfield  {author} {\bibinfo {author} {\bibfnamefont {W.}~\bibnamefont {Chen}}, \bibinfo {author} {\bibfnamefont {X.}~\bibnamefont {Huang}}, \bibinfo {author} {\bibfnamefont {D.~V.}\ \bibnamefont {Semenok}}, \bibinfo {author} {\bibfnamefont {S.}~\bibnamefont {Chen}}, \bibinfo {author} {\bibfnamefont {D.}~\bibnamefont {Zhou}}, \bibinfo {author} {\bibfnamefont {K.}~\bibnamefont {Zhang}}, \bibinfo {author} {\bibfnamefont {A.~R.}\ \bibnamefont {Oganov}},\ and\ \bibinfo {author} {\bibfnamefont {T.}~\bibnamefont {Cui}},\ }\bibfield  {title} {\bibinfo {title} {Enhancement of superconducting properties in the {La--Ce--H} system at moderate pressures},\ }\href {https://doi.org/10.1038/s41467-023-38254-6} {\bibfield  {journal} {\bibinfo  {journal} {Nat. Commun.}\ }\textbf {\bibinfo {volume} {14}},\ \bibinfo {pages} {2660} (\bibinfo {year} {2023})}\BibitemShut {NoStop}%
\bibitem [{\citenamefont {Jiang}\ \emph {et~al.}(2024)\citenamefont {Jiang}, \citenamefont {Zhang}, \citenamefont {Song}, \citenamefont {Ma}, \citenamefont {Sun}, \citenamefont {Miao}, \citenamefont {Cui},\ and\ \citenamefont {Duan}}]{jiang2024ternary}%
  \BibitemOpen
  \bibfield  {author} {\bibinfo {author} {\bibfnamefont {Q.}~\bibnamefont {Jiang}}, \bibinfo {author} {\bibfnamefont {Z.}~\bibnamefont {Zhang}}, \bibinfo {author} {\bibfnamefont {H.}~\bibnamefont {Song}}, \bibinfo {author} {\bibfnamefont {Y.}~\bibnamefont {Ma}}, \bibinfo {author} {\bibfnamefont {Y.}~\bibnamefont {Sun}}, \bibinfo {author} {\bibfnamefont {M.}~\bibnamefont {Miao}}, \bibinfo {author} {\bibfnamefont {T.}~\bibnamefont {Cui}},\ and\ \bibinfo {author} {\bibfnamefont {D.}~\bibnamefont {Duan}},\ }\bibfield  {title} {\bibinfo {title} {Ternary superconducting hydrides stabilized via {Th} and {Ce} elements at mild pressures},\ }\href {https://doi.org/10.1016/j.fmre.2022.11.010} {\bibfield  {journal} {\bibinfo  {journal} {Fundam. Res.}\ }\textbf {\bibinfo {volume} {4}},\ \bibinfo {pages} {550} (\bibinfo {year} {2024})}\BibitemShut {NoStop}%
\bibitem [{\citenamefont {Di~Cataldo}\ \emph {et~al.}(2021)\citenamefont {Di~Cataldo}, \citenamefont {Heil}, \citenamefont {von~der Linden},\ and\ \citenamefont {Boeri}}]{di2021bh}%
  \BibitemOpen
  \bibfield  {author} {\bibinfo {author} {\bibfnamefont {S.}~\bibnamefont {Di~Cataldo}}, \bibinfo {author} {\bibfnamefont {C.}~\bibnamefont {Heil}}, \bibinfo {author} {\bibfnamefont {W.}~\bibnamefont {von~der Linden}},\ and\ \bibinfo {author} {\bibfnamefont {L.}~\bibnamefont {Boeri}},\ }\bibfield  {title} {\bibinfo {title} {{LaBH$_8$}: Toward high-{$T_c$} low-pressure superconductivity in ternary superhydrides},\ }\href {https://doi.org/10.1103/PhysRevB.104.L020511} {\bibfield  {journal} {\bibinfo  {journal} {Phys. Rev. B}\ }\textbf {\bibinfo {volume} {104}},\ \bibinfo {pages} {L020511} (\bibinfo {year} {2021})}\BibitemShut {NoStop}%
\bibitem [{\citenamefont {Liang}\ \emph {et~al.}(2021)\citenamefont {Liang}, \citenamefont {Bergara}, \citenamefont {Wei}, \citenamefont {Song}, \citenamefont {Wang}, \citenamefont {Sun}, \citenamefont {Liu}, \citenamefont {Hemley}, \citenamefont {Wang}, \citenamefont {Gao} \emph {et~al.}}]{liang2021prediction}%
  \BibitemOpen
  \bibfield  {author} {\bibinfo {author} {\bibfnamefont {X.}~\bibnamefont {Liang}}, \bibinfo {author} {\bibfnamefont {A.}~\bibnamefont {Bergara}}, \bibinfo {author} {\bibfnamefont {X.}~\bibnamefont {Wei}}, \bibinfo {author} {\bibfnamefont {X.}~\bibnamefont {Song}}, \bibinfo {author} {\bibfnamefont {L.}~\bibnamefont {Wang}}, \bibinfo {author} {\bibfnamefont {R.}~\bibnamefont {Sun}}, \bibinfo {author} {\bibfnamefont {H.}~\bibnamefont {Liu}}, \bibinfo {author} {\bibfnamefont {R.~J.}\ \bibnamefont {Hemley}}, \bibinfo {author} {\bibfnamefont {L.}~\bibnamefont {Wang}}, \bibinfo {author} {\bibfnamefont {G.}~\bibnamefont {Gao}}, \emph {et~al.},\ }\bibfield  {title} {\bibinfo {title} {Prediction of high-{$T_c$} superconductivity in ternary lanthanum borohydrides},\ }\href {https://doi.org/10.1103/PhysRevB.104.134501} {\bibfield  {journal} {\bibinfo  {journal} {Phys. Rev. B}\ }\textbf {\bibinfo {volume} {104}},\ \bibinfo {pages} {134501} (\bibinfo {year} {2021})}\BibitemShut {NoStop}%
\bibitem [{\citenamefont {Lucrezi}\ \emph {et~al.}(2022)\citenamefont {Lucrezi}, \citenamefont {Di~Cataldo}, \citenamefont {von~der Linden}, \citenamefont {Boeri},\ and\ \citenamefont {Heil}}]{lucrezi2022silico}%
  \BibitemOpen
  \bibfield  {author} {\bibinfo {author} {\bibfnamefont {R.}~\bibnamefont {Lucrezi}}, \bibinfo {author} {\bibfnamefont {S.}~\bibnamefont {Di~Cataldo}}, \bibinfo {author} {\bibfnamefont {W.}~\bibnamefont {von~der Linden}}, \bibinfo {author} {\bibfnamefont {L.}~\bibnamefont {Boeri}},\ and\ \bibinfo {author} {\bibfnamefont {C.}~\bibnamefont {Heil}},\ }\bibfield  {title} {\bibinfo {title} {In-silico synthesis of lowest-pressure high-{$T_c$} ternary superhydrides},\ }\href {https://doi.org/10.1038/s41524-022-00801-y} {\bibfield  {journal} {\bibinfo  {journal} {npj Comput. Mater.}\ }\textbf {\bibinfo {volume} {8}},\ \bibinfo {pages} {119} (\bibinfo {year} {2022})}\BibitemShut {NoStop}%
\bibitem [{\citenamefont {Zhang}\ \emph {et~al.}(2022)\citenamefont {Zhang}, \citenamefont {Cui}, \citenamefont {Hutcheon}, \citenamefont {Shipley}, \citenamefont {Song}, \citenamefont {Du}, \citenamefont {Kresin}, \citenamefont {Duan}, \citenamefont {Pickard},\ and\ \citenamefont {Yao}}]{zhang2022design}%
  \BibitemOpen
  \bibfield  {author} {\bibinfo {author} {\bibfnamefont {Z.}~\bibnamefont {Zhang}}, \bibinfo {author} {\bibfnamefont {T.}~\bibnamefont {Cui}}, \bibinfo {author} {\bibfnamefont {M.~J.}\ \bibnamefont {Hutcheon}}, \bibinfo {author} {\bibfnamefont {A.~M.}\ \bibnamefont {Shipley}}, \bibinfo {author} {\bibfnamefont {H.}~\bibnamefont {Song}}, \bibinfo {author} {\bibfnamefont {M.}~\bibnamefont {Du}}, \bibinfo {author} {\bibfnamefont {V.~Z.}\ \bibnamefont {Kresin}}, \bibinfo {author} {\bibfnamefont {D.}~\bibnamefont {Duan}}, \bibinfo {author} {\bibfnamefont {C.~J.}\ \bibnamefont {Pickard}},\ and\ \bibinfo {author} {\bibfnamefont {Y.}~\bibnamefont {Yao}},\ }\bibfield  {title} {\bibinfo {title} {Design principles for high-temperature superconductors with a hydrogen-based alloy backbone at moderate pressure},\ }\href {https://doi.org/10.1103/PhysRevLett.128.047001} {\bibfield  {journal} {\bibinfo  {journal} {Phys. Rev. Lett.}\ }\textbf {\bibinfo {volume} {128}},\ \bibinfo {pages} {047001} (\bibinfo {year}
  {2022})}\BibitemShut {NoStop}%
\bibitem [{\citenamefont {Sanna}\ \emph {et~al.}(2024)\citenamefont {Sanna}, \citenamefont {Cerqueira}, \citenamefont {Fang}, \citenamefont {Errea}, \citenamefont {Ludwig},\ and\ \citenamefont {Marques}}]{sanna2024prediction}%
  \BibitemOpen
  \bibfield  {author} {\bibinfo {author} {\bibfnamefont {A.}~\bibnamefont {Sanna}}, \bibinfo {author} {\bibfnamefont {T.~F.~T.}\ \bibnamefont {Cerqueira}}, \bibinfo {author} {\bibfnamefont {Y.-W.}\ \bibnamefont {Fang}}, \bibinfo {author} {\bibfnamefont {I.}~\bibnamefont {Errea}}, \bibinfo {author} {\bibfnamefont {A.}~\bibnamefont {Ludwig}},\ and\ \bibinfo {author} {\bibfnamefont {M.~A.~L.}\ \bibnamefont {Marques}},\ }\bibfield  {title} {\bibinfo {title} {Prediction of ambient pressure conventional superconductivity above 80 {K} in hydride compounds},\ }\href {https://doi.org/10.1038/s41524-024-01214-9} {\bibfield  {journal} {\bibinfo  {journal} {npj Comput. Mater.}\ }\textbf {\bibinfo {volume} {10}},\ \bibinfo {pages} {44} (\bibinfo {year} {2024})}\BibitemShut {NoStop}%
\bibitem [{\citenamefont {Cerqueira}\ \emph {et~al.}(2024)\citenamefont {Cerqueira}, \citenamefont {Fang}, \citenamefont {Errea}, \citenamefont {Sanna},\ and\ \citenamefont {Marques}}]{cerqueira2024searching}%
  \BibitemOpen
  \bibfield  {author} {\bibinfo {author} {\bibfnamefont {T.~F.~T.}\ \bibnamefont {Cerqueira}}, \bibinfo {author} {\bibfnamefont {Y.-W.}\ \bibnamefont {Fang}}, \bibinfo {author} {\bibfnamefont {I.}~\bibnamefont {Errea}}, \bibinfo {author} {\bibfnamefont {A.}~\bibnamefont {Sanna}},\ and\ \bibinfo {author} {\bibfnamefont {M.~A.~L.}\ \bibnamefont {Marques}},\ }\bibfield  {title} {\bibinfo {title} {Searching materials space for hydride superconductors at ambient pressure},\ }\href {https://doi.org/10.1002/adfm.202404043} {\bibfield  {journal} {\bibinfo  {journal} {Adv. Funct. Mater.}\ }\textbf {\bibinfo {volume} {34}},\ \bibinfo {pages} {2404043} (\bibinfo {year} {2024})}\BibitemShut {NoStop}%
\bibitem [{\citenamefont {Dolui}\ \emph {et~al.}(2024)\citenamefont {Dolui}, \citenamefont {Conway}, \citenamefont {Heil}, \citenamefont {Strobel}, \citenamefont {Prasankumar},\ and\ \citenamefont {Pickard}}]{dolui2024feasible}%
  \BibitemOpen
  \bibfield  {author} {\bibinfo {author} {\bibfnamefont {K.}~\bibnamefont {Dolui}}, \bibinfo {author} {\bibfnamefont {L.~J.}\ \bibnamefont {Conway}}, \bibinfo {author} {\bibfnamefont {C.}~\bibnamefont {Heil}}, \bibinfo {author} {\bibfnamefont {T.~A.}\ \bibnamefont {Strobel}}, \bibinfo {author} {\bibfnamefont {R.~P.}\ \bibnamefont {Prasankumar}},\ and\ \bibinfo {author} {\bibfnamefont {C.~J.}\ \bibnamefont {Pickard}},\ }\bibfield  {title} {\bibinfo {title} {Feasible route to high-temperature ambient-pressure hydride superconductivity},\ }\href {https://doi.org/10.1103/PhysRevLett.132.166001} {\bibfield  {journal} {\bibinfo  {journal} {Phys. Rev. Lett.}\ }\textbf {\bibinfo {volume} {132}},\ \bibinfo {pages} {166001} (\bibinfo {year} {2024})}\BibitemShut {NoStop}%
\bibitem [{\citenamefont {Ouyang}\ \emph {et~al.}(2025)\citenamefont {Ouyang}, \citenamefont {Yao}, \citenamefont {Han}, \citenamefont {Guo}, \citenamefont {Gao},\ and\ \citenamefont {Lu}}]{ouyang2025strong}%
  \BibitemOpen
  \bibfield  {author} {\bibinfo {author} {\bibfnamefont {Z.}~\bibnamefont {Ouyang}}, \bibinfo {author} {\bibfnamefont {B.-W.}\ \bibnamefont {Yao}}, \bibinfo {author} {\bibfnamefont {X.-Q.}\ \bibnamefont {Han}}, \bibinfo {author} {\bibfnamefont {P.-J.}\ \bibnamefont {Guo}}, \bibinfo {author} {\bibfnamefont {Z.-F.}\ \bibnamefont {Gao}},\ and\ \bibinfo {author} {\bibfnamefont {Z.-Y.}\ \bibnamefont {Lu}},\ }\bibfield  {title} {\bibinfo {title} {High-temperature superconductivity in {Li$_2$AuH$_6$} mediated by strong electron-phonon coupling under ambient pressure},\ }\href {https://doi.org/10.1103/PhysRevB.111.L140501} {\bibfield  {journal} {\bibinfo  {journal} {Phys. Rev. B}\ }\textbf {\bibinfo {volume} {111}},\ \bibinfo {pages} {L140501} (\bibinfo {year} {2025})}\BibitemShut {NoStop}%
\bibitem [{\citenamefont {Han}\ \emph {et~al.}(2025{\natexlab{a}})\citenamefont {Han}, \citenamefont {Ouyang}, \citenamefont {Guo}, \citenamefont {Sun}, \citenamefont {Gao},\ and\ \citenamefont {Lu}}]{xiao2025invdesflow}%
  \BibitemOpen
  \bibfield  {author} {\bibinfo {author} {\bibfnamefont {X.-Q.}\ \bibnamefont {Han}}, \bibinfo {author} {\bibfnamefont {Z.}~\bibnamefont {Ouyang}}, \bibinfo {author} {\bibfnamefont {P.-J.}\ \bibnamefont {Guo}}, \bibinfo {author} {\bibfnamefont {H.}~\bibnamefont {Sun}}, \bibinfo {author} {\bibfnamefont {Z.-F.}\ \bibnamefont {Gao}},\ and\ \bibinfo {author} {\bibfnamefont {Z.-Y.}\ \bibnamefont {Lu}},\ }\bibfield  {title} {\bibinfo {title} {{InvDesFlow}: An {AI}-driven materials inverse design workflow to explore possible high-temperature superconductors},\ }\href {https://doi.org/10.1088/0256-307X/42/4/047301} {\bibfield  {journal} {\bibinfo  {journal} {Chin. Phys. Lett.}\ }\textbf {\bibinfo {volume} {42}},\ \bibinfo {pages} {047301} (\bibinfo {year} {2025}{\natexlab{a}})}\BibitemShut {NoStop}%
\bibitem [{\citenamefont {Han}\ \emph {et~al.}(2025{\natexlab{b}})\citenamefont {Han}, \citenamefont {Guo}, \citenamefont {Gao}, \citenamefont {Sun},\ and\ \citenamefont {Lu}}]{InvDesFlow-AL}%
  \BibitemOpen
  \bibfield  {author} {\bibinfo {author} {\bibfnamefont {X.-Q.}\ \bibnamefont {Han}}, \bibinfo {author} {\bibfnamefont {P.-J.}\ \bibnamefont {Guo}}, \bibinfo {author} {\bibfnamefont {Z.-F.}\ \bibnamefont {Gao}}, \bibinfo {author} {\bibfnamefont {H.}~\bibnamefont {Sun}},\ and\ \bibinfo {author} {\bibfnamefont {Z.-Y.}\ \bibnamefont {Lu}},\ }\href {https://doi.org/10.48550/arXiv.2505.09203} {\bibinfo {title} {{InvDesFlow-AL}: {Active Learning}-based workflow for inverse design of functional materials}} (\bibinfo {year} {2025}{\natexlab{b}}),\ \bibinfo {note} {preprint submitted to \textit{Phys. Rev. Mater.}},\ \Eprint {https://arxiv.org/abs/2505.09203} {arXiv:2505.09203 [cond-mat.mtrl-sci]} \BibitemShut {NoStop}%
\bibitem [{SM()}]{SM}%
  \BibitemOpen
  \href@noop {} {\bibinfo {title} {See {Supplemental} {Material} at xxx for the detailed {AI} method and more calculated results of {Na-Cu-H} and {Na-Ag-H} systems, which includes {Refs}.~\cite{song2020score,han2024aidriven,han2024surveygeometricgraphneural,jiao2024crystal,jiao2024space,YE2024100003,Antunes2024,cerqueira2024searching,sanna2024prediction,gnome,mattergen,xiao2025invdesflow,dolui2024feasible}}}\BibitemShut {NoStop}%
\bibitem [{cod()}]{code}%
  \BibitemOpen
  \href {https://github.com/xqh19970407/InvDesFlow} {\bibinfo  {journal} {https://github.com/xqh19970407/InvDesFlow}\ }\BibitemShut {NoStop}%
\bibitem [{\citenamefont {Hohenberg}\ and\ \citenamefont {Kohn}(1964)}]{hohenberg1964inhomogeneous}%
  \BibitemOpen
\bibfield  {journal} {  }\bibfield  {author} {\bibinfo {author} {\bibfnamefont {P.}~\bibnamefont {Hohenberg}}\ and\ \bibinfo {author} {\bibfnamefont {W.}~\bibnamefont {Kohn}},\ }\bibfield  {title} {\bibinfo {title} {Inhomogeneous electron gas},\ }\href {https://doi.org/10.1103/PhysRev.136.B864} {\bibfield  {journal} {\bibinfo  {journal} {Phys. Rev.}\ }\textbf {\bibinfo {volume} {136}},\ \bibinfo {pages} {B864} (\bibinfo {year} {1964})}\BibitemShut {NoStop}%
\bibitem [{\citenamefont {Kohn}\ and\ \citenamefont {Sham}(1965)}]{kohn1965self}%
  \BibitemOpen
  \bibfield  {author} {\bibinfo {author} {\bibfnamefont {W.}~\bibnamefont {Kohn}}\ and\ \bibinfo {author} {\bibfnamefont {L.~J.}\ \bibnamefont {Sham}},\ }\bibfield  {title} {\bibinfo {title} {Self-consistent equations including exchange and correlation effects},\ }\href {https://doi.org/10.1103/PhysRev.140.A1133} {\bibfield  {journal} {\bibinfo  {journal} {Phys. Rev.}\ }\textbf {\bibinfo {volume} {140}},\ \bibinfo {pages} {A1133} (\bibinfo {year} {1965})}\BibitemShut {NoStop}%
\bibitem [{\citenamefont {Giannozzi}\ \emph {et~al.}(2009)\citenamefont {Giannozzi}, \citenamefont {Baroni}, \citenamefont {Bonini}, \citenamefont {Calandra}, \citenamefont {Car}, \citenamefont {Cavazzoni}, \citenamefont {Ceresoli}, \citenamefont {Chiarotti}, \citenamefont {Cococcioni}, \citenamefont {Dabo} \emph {et~al.}}]{giannozzi2009quantum}%
  \BibitemOpen
  \bibfield  {author} {\bibinfo {author} {\bibfnamefont {P.}~\bibnamefont {Giannozzi}}, \bibinfo {author} {\bibfnamefont {S.}~\bibnamefont {Baroni}}, \bibinfo {author} {\bibfnamefont {N.}~\bibnamefont {Bonini}}, \bibinfo {author} {\bibfnamefont {M.}~\bibnamefont {Calandra}}, \bibinfo {author} {\bibfnamefont {R.}~\bibnamefont {Car}}, \bibinfo {author} {\bibfnamefont {C.}~\bibnamefont {Cavazzoni}}, \bibinfo {author} {\bibfnamefont {D.}~\bibnamefont {Ceresoli}}, \bibinfo {author} {\bibfnamefont {G.~L.}\ \bibnamefont {Chiarotti}}, \bibinfo {author} {\bibfnamefont {M.}~\bibnamefont {Cococcioni}}, \bibinfo {author} {\bibfnamefont {I.}~\bibnamefont {Dabo}}, \emph {et~al.},\ }\bibfield  {title} {\bibinfo {title} {{QUANTUM ESPRESSO}: a modular and open-source software project for quantum simulations of materials},\ }\href {https://doi.org/10.1088/0953-8984/21/39/395502} {\bibfield  {journal} {\bibinfo  {journal} {J. Phys.: Condens. Matter}\ }\textbf {\bibinfo {volume} {21}},\ \bibinfo {pages} {395502} (\bibinfo {year}
  {2009})}\BibitemShut {NoStop}%
\bibitem [{\citenamefont {Perdew}\ \emph {et~al.}(1996)\citenamefont {Perdew}, \citenamefont {Burke},\ and\ \citenamefont {Ernzerhof}}]{perdew1996generalized}%
  \BibitemOpen
  \bibfield  {author} {\bibinfo {author} {\bibfnamefont {J.~P.}\ \bibnamefont {Perdew}}, \bibinfo {author} {\bibfnamefont {K.}~\bibnamefont {Burke}},\ and\ \bibinfo {author} {\bibfnamefont {M.}~\bibnamefont {Ernzerhof}},\ }\bibfield  {title} {\bibinfo {title} {Generalized gradient approximation made simple},\ }\href {https://doi.org/10.1103/PhysRevLett.77.3865} {\bibfield  {journal} {\bibinfo  {journal} {Phys. Rev. Lett.}\ }\textbf {\bibinfo {volume} {77}},\ \bibinfo {pages} {3865} (\bibinfo {year} {1996})}\BibitemShut {NoStop}%
\bibitem [{\citenamefont {Hamann}(2013)}]{hamann2013optimized}%
  \BibitemOpen
  \bibfield  {author} {\bibinfo {author} {\bibfnamefont {D.~R.}\ \bibnamefont {Hamann}},\ }\bibfield  {title} {\bibinfo {title} {Optimized norm-conserving {Vanderbilt} pseudopotentials},\ }\href {https://doi.org/10.1103/PhysRevB.88.085117} {\bibfield  {journal} {\bibinfo  {journal} {Phys. Rev. B}\ }\textbf {\bibinfo {volume} {88}},\ \bibinfo {pages} {085117} (\bibinfo {year} {2013})}\BibitemShut {NoStop}%
\bibitem [{\citenamefont {Methfessel}\ and\ \citenamefont {Paxton}(1989)}]{methfessel1989high}%
  \BibitemOpen
  \bibfield  {author} {\bibinfo {author} {\bibfnamefont {M.}~\bibnamefont {Methfessel}}\ and\ \bibinfo {author} {\bibfnamefont {A.~T.}\ \bibnamefont {Paxton}},\ }\bibfield  {title} {\bibinfo {title} {High-precision sampling for {Brillouin}-zone integration in metals},\ }\href {https://doi.org/10.1103/PhysRevB.40.3616} {\bibfield  {journal} {\bibinfo  {journal} {Phys. Rev. B}\ }\textbf {\bibinfo {volume} {40}},\ \bibinfo {pages} {3616} (\bibinfo {year} {1989})}\BibitemShut {NoStop}%
\bibitem [{\citenamefont {Baroni}\ \emph {et~al.}(2001)\citenamefont {Baroni}, \citenamefont {de~Gironcoli}, \citenamefont {Dal~Corso},\ and\ \citenamefont {Giannozzi}}]{baroni2001phonons}%
  \BibitemOpen
  \bibfield  {author} {\bibinfo {author} {\bibfnamefont {S.}~\bibnamefont {Baroni}}, \bibinfo {author} {\bibfnamefont {S.}~\bibnamefont {de~Gironcoli}}, \bibinfo {author} {\bibfnamefont {A.}~\bibnamefont {Dal~Corso}},\ and\ \bibinfo {author} {\bibfnamefont {P.}~\bibnamefont {Giannozzi}},\ }\bibfield  {title} {\bibinfo {title} {Phonons and related crystal properties from density-functional perturbation theory},\ }\href {https://doi.org/10.1103/RevModPhys.73.515} {\bibfield  {journal} {\bibinfo  {journal} {Rev. Mod. Phys.}\ }\textbf {\bibinfo {volume} {73}},\ \bibinfo {pages} {515} (\bibinfo {year} {2001})}\BibitemShut {NoStop}%
\bibitem [{\citenamefont {Pizzi}\ \emph {et~al.}(2020)\citenamefont {Pizzi}, \citenamefont {Vitale}, \citenamefont {Arita}, \citenamefont {Bl\"ugel}, \citenamefont {Freimuth}, \citenamefont {G\'eranton}, \citenamefont {Gibertini}, \citenamefont {Gresch}, \citenamefont {Johnson}, \citenamefont {Koretsune} \emph {et~al.}}]{pizzi2020wannier90}%
  \BibitemOpen
  \bibfield  {author} {\bibinfo {author} {\bibfnamefont {G.}~\bibnamefont {Pizzi}}, \bibinfo {author} {\bibfnamefont {V.}~\bibnamefont {Vitale}}, \bibinfo {author} {\bibfnamefont {R.}~\bibnamefont {Arita}}, \bibinfo {author} {\bibfnamefont {S.}~\bibnamefont {Bl\"ugel}}, \bibinfo {author} {\bibfnamefont {F.}~\bibnamefont {Freimuth}}, \bibinfo {author} {\bibfnamefont {G.}~\bibnamefont {G\'eranton}}, \bibinfo {author} {\bibfnamefont {M.}~\bibnamefont {Gibertini}}, \bibinfo {author} {\bibfnamefont {D.}~\bibnamefont {Gresch}}, \bibinfo {author} {\bibfnamefont {C.}~\bibnamefont {Johnson}}, \bibinfo {author} {\bibfnamefont {T.}~\bibnamefont {Koretsune}}, \emph {et~al.},\ }\bibfield  {title} {\bibinfo {title} {{Wannier90} as a community code: new features and applications},\ }\href {https://doi.org/10.1088/1361-648X/ab51ff} {\bibfield  {journal} {\bibinfo  {journal} {J. Phys.: Condens. Matter}\ }\textbf {\bibinfo {volume} {32}},\ \bibinfo {pages} {165902} (\bibinfo {year} {2020})}\BibitemShut {NoStop}%
\bibitem [{\citenamefont {Ponc\'e}\ \emph {et~al.}(2016)\citenamefont {Ponc\'e}, \citenamefont {Margine}, \citenamefont {Verdi},\ and\ \citenamefont {Giustino}}]{ponce2016epw}%
  \BibitemOpen
  \bibfield  {author} {\bibinfo {author} {\bibfnamefont {S.}~\bibnamefont {Ponc\'e}}, \bibinfo {author} {\bibfnamefont {E.~R.}\ \bibnamefont {Margine}}, \bibinfo {author} {\bibfnamefont {C.}~\bibnamefont {Verdi}},\ and\ \bibinfo {author} {\bibfnamefont {F.}~\bibnamefont {Giustino}},\ }\bibfield  {title} {\bibinfo {title} {{EPW}: {E}lectron--phonon coupling, transport and superconducting properties using maximally localized {W}annier functions},\ }\href {https://doi.org/10.1016/j.cpc.2016.07.028} {\bibfield  {journal} {\bibinfo  {journal} {Comput. Phys. Commun.}\ }\textbf {\bibinfo {volume} {209}},\ \bibinfo {pages} {116} (\bibinfo {year} {2016})}\BibitemShut {NoStop}%
\bibitem [{\citenamefont {Choi}\ \emph {et~al.}(2003)\citenamefont {Choi}, \citenamefont {Cohen},\ and\ \citenamefont {Louie}}]{choi2003anisotropic}%
  \BibitemOpen
  \bibfield  {author} {\bibinfo {author} {\bibfnamefont {H.~J.}\ \bibnamefont {Choi}}, \bibinfo {author} {\bibfnamefont {M.~L.}\ \bibnamefont {Cohen}},\ and\ \bibinfo {author} {\bibfnamefont {S.~G.}\ \bibnamefont {Louie}},\ }\bibfield  {title} {\bibinfo {title} {Anisotropic {E}liashberg theory of {MgB}$_2$: {$T_c$}, isotope effects, superconducting energy gaps, quasiparticles, and specific heat},\ }\href {https://doi.org/10.1016/S0921-4534(02)02345-6} {\bibfield  {journal} {\bibinfo  {journal} {Physica C}\ }\textbf {\bibinfo {volume} {385}},\ \bibinfo {pages} {66} (\bibinfo {year} {2003})}\BibitemShut {NoStop}%
\bibitem [{\citenamefont {Margine}\ and\ \citenamefont {Giustino}(2013)}]{margine2013anisotropic}%
  \BibitemOpen
  \bibfield  {author} {\bibinfo {author} {\bibfnamefont {E.~R.}\ \bibnamefont {Margine}}\ and\ \bibinfo {author} {\bibfnamefont {F.}~\bibnamefont {Giustino}},\ }\bibfield  {title} {\bibinfo {title} {Anisotropic {M}igdal-{E}liashberg theory using {W}annier functions},\ }\href {https://doi.org/10.1103/PhysRevB.87.024505} {\bibfield  {journal} {\bibinfo  {journal} {Phys. Rev. B}\ }\textbf {\bibinfo {volume} {87}},\ \bibinfo {pages} {024505} (\bibinfo {year} {2013})}\BibitemShut {NoStop}%
\bibitem [{\citenamefont {Scarselli}\ \emph {et~al.}(2008)\citenamefont {Scarselli}, \citenamefont {Gori}, \citenamefont {Tsoi}, \citenamefont {Hagenbuchner},\ and\ \citenamefont {Monfardini}}]{scarselli2008graph}%
  \BibitemOpen
  \bibfield  {author} {\bibinfo {author} {\bibfnamefont {F.}~\bibnamefont {Scarselli}}, \bibinfo {author} {\bibfnamefont {M.}~\bibnamefont {Gori}}, \bibinfo {author} {\bibfnamefont {A.~C.}\ \bibnamefont {Tsoi}}, \bibinfo {author} {\bibfnamefont {M.}~\bibnamefont {Hagenbuchner}},\ and\ \bibinfo {author} {\bibfnamefont {G.}~\bibnamefont {Monfardini}},\ }\bibfield  {title} {\bibinfo {title} {The graph neural network model},\ }\href {https://doi.org/10.1109/TNN.2008.2005605} {\bibfield  {journal} {\bibinfo  {journal} {IEEE Trans. Neural Netw.}\ }\textbf {\bibinfo {volume} {20}},\ \bibinfo {pages} {61} (\bibinfo {year} {2008})}\BibitemShut {NoStop}%
\bibitem [{\citenamefont {Song}\ \emph {et~al.}(2021)\citenamefont {Song}, \citenamefont {Sohl-Dickstein}, \citenamefont {Kingma}, \citenamefont {Kumar}, \citenamefont {Ermon},\ and\ \citenamefont {Poole}}]{song2020score}%
  \BibitemOpen
  \bibfield  {author} {\bibinfo {author} {\bibfnamefont {Y.}~\bibnamefont {Song}}, \bibinfo {author} {\bibfnamefont {J.}~\bibnamefont {Sohl-Dickstein}}, \bibinfo {author} {\bibfnamefont {D.~P.}\ \bibnamefont {Kingma}}, \bibinfo {author} {\bibfnamefont {A.}~\bibnamefont {Kumar}}, \bibinfo {author} {\bibfnamefont {S.}~\bibnamefont {Ermon}},\ and\ \bibinfo {author} {\bibfnamefont {B.}~\bibnamefont {Poole}},\ }\href {https://arxiv.org/abs/2011.13456} {\bibinfo {title} {Score-based generative modeling through stochastic differential equations}} (\bibinfo {year} {2021}),\ \Eprint {https://arxiv.org/abs/2011.13456} {arXiv:2011.13456 [cs.LG]} \BibitemShut {NoStop}%
\bibitem [{\citenamefont {Han}\ \emph {et~al.}(2025{\natexlab{c}})\citenamefont {Han}, \citenamefont {Wang}, \citenamefont {Xu}, \citenamefont {Feng}, \citenamefont {Yao}, \citenamefont {Guo}, \citenamefont {Gao},\ and\ \citenamefont {Lu}}]{han2024aidriven}%
  \BibitemOpen
  \bibfield  {author} {\bibinfo {author} {\bibfnamefont {X.-Q.}\ \bibnamefont {Han}}, \bibinfo {author} {\bibfnamefont {X.-D.}\ \bibnamefont {Wang}}, \bibinfo {author} {\bibfnamefont {M.-Y.}\ \bibnamefont {Xu}}, \bibinfo {author} {\bibfnamefont {Z.}~\bibnamefont {Feng}}, \bibinfo {author} {\bibfnamefont {B.-W.}\ \bibnamefont {Yao}}, \bibinfo {author} {\bibfnamefont {P.-J.}\ \bibnamefont {Guo}}, \bibinfo {author} {\bibfnamefont {Z.-F.}\ \bibnamefont {Gao}},\ and\ \bibinfo {author} {\bibfnamefont {Z.-Y.}\ \bibnamefont {Lu}},\ }\bibfield  {title} {\bibinfo {title} {{AI}-driven inverse design of materials: Past, present, and future},\ }\href {https://doi.org/10.1088/0256-307X/42/2/027403} {\bibfield  {journal} {\bibinfo  {journal} {Chin. Phys. Lett.}\ }\textbf {\bibinfo {volume} {42}},\ \bibinfo {pages} {027403} (\bibinfo {year} {2025}{\natexlab{c}})}\BibitemShut {NoStop}%
\bibitem [{\citenamefont {Han}\ \emph {et~al.}(2024)\citenamefont {Han}, \citenamefont {Cen}, \citenamefont {Wu}, \citenamefont {Li}, \citenamefont {Kong}, \citenamefont {Jiao}, \citenamefont {Yu}, \citenamefont {Xu}, \citenamefont {Wu}, \citenamefont {Wang}, \citenamefont {Xu}, \citenamefont {Wei}, \citenamefont {Liu}, \citenamefont {Rong},\ and\ \citenamefont {Huang}}]{han2024surveygeometricgraphneural}%
  \BibitemOpen
  \bibfield  {author} {\bibinfo {author} {\bibfnamefont {J.}~\bibnamefont {Han}}, \bibinfo {author} {\bibfnamefont {J.}~\bibnamefont {Cen}}, \bibinfo {author} {\bibfnamefont {L.}~\bibnamefont {Wu}}, \bibinfo {author} {\bibfnamefont {Z.}~\bibnamefont {Li}}, \bibinfo {author} {\bibfnamefont {X.}~\bibnamefont {Kong}}, \bibinfo {author} {\bibfnamefont {R.}~\bibnamefont {Jiao}}, \bibinfo {author} {\bibfnamefont {Z.}~\bibnamefont {Yu}}, \bibinfo {author} {\bibfnamefont {T.}~\bibnamefont {Xu}}, \bibinfo {author} {\bibfnamefont {F.}~\bibnamefont {Wu}}, \bibinfo {author} {\bibfnamefont {Z.}~\bibnamefont {Wang}}, \bibinfo {author} {\bibfnamefont {H.}~\bibnamefont {Xu}}, \bibinfo {author} {\bibfnamefont {Z.}~\bibnamefont {Wei}}, \bibinfo {author} {\bibfnamefont {Y.}~\bibnamefont {Liu}}, \bibinfo {author} {\bibfnamefont {Y.}~\bibnamefont {Rong}},\ and\ \bibinfo {author} {\bibfnamefont {W.}~\bibnamefont {Huang}},\ }\href {https://arxiv.org/abs/2403.00485} {\bibinfo {title} {A survey of geometric graph neural networks:
  Data structures, models and applications}} (\bibinfo {year} {2024}),\ \Eprint {https://arxiv.org/abs/2403.00485} {arXiv:2403.00485 [cs.LG]} \BibitemShut {NoStop}%
\bibitem [{\citenamefont {Jiao}\ \emph {et~al.}(2023)\citenamefont {Jiao}, \citenamefont {Huang}, \citenamefont {Lin}, \citenamefont {Han}, \citenamefont {Chen}, \citenamefont {Lu},\ and\ \citenamefont {Liu}}]{jiao2024crystal}%
  \BibitemOpen
  \bibfield  {author} {\bibinfo {author} {\bibfnamefont {R.}~\bibnamefont {Jiao}}, \bibinfo {author} {\bibfnamefont {W.}~\bibnamefont {Huang}}, \bibinfo {author} {\bibfnamefont {P.}~\bibnamefont {Lin}}, \bibinfo {author} {\bibfnamefont {J.}~\bibnamefont {Han}}, \bibinfo {author} {\bibfnamefont {P.}~\bibnamefont {Chen}}, \bibinfo {author} {\bibfnamefont {Y.}~\bibnamefont {Lu}},\ and\ \bibinfo {author} {\bibfnamefont {Y.}~\bibnamefont {Liu}},\ }\bibfield  {title} {\bibinfo {title} {Crystal structure prediction by joint equivariant diffusion},\ }\href {https://proceedings.neurips.cc/paper_files/paper/2023/file/38b787fc530d0b31825827e2cc306656-Paper-Conference.pdf} {\bibfield  {journal} {\bibinfo  {journal} {NeurIPS}\ }\textbf {\bibinfo {volume} {36}},\ \bibinfo {pages} {17464} (\bibinfo {year} {2023})}\BibitemShut {NoStop}%
\bibitem [{\citenamefont {Jiao}\ \emph {et~al.}(2024)\citenamefont {Jiao}, \citenamefont {Huang}, \citenamefont {Liu}, \citenamefont {Zhao},\ and\ \citenamefont {Liu}}]{jiao2024space}%
  \BibitemOpen
  \bibfield  {author} {\bibinfo {author} {\bibfnamefont {R.}~\bibnamefont {Jiao}}, \bibinfo {author} {\bibfnamefont {W.}~\bibnamefont {Huang}}, \bibinfo {author} {\bibfnamefont {Y.}~\bibnamefont {Liu}}, \bibinfo {author} {\bibfnamefont {D.}~\bibnamefont {Zhao}},\ and\ \bibinfo {author} {\bibfnamefont {Y.}~\bibnamefont {Liu}},\ }\href {https://doi.org/https://arxiv.org/abs/2402.03992} {\bibinfo {title} {Space group constrained crystal generation}} (\bibinfo {year} {2024}),\ \Eprint {https://arxiv.org/abs/2402.03992} {arXiv:2402.03992 [cs.LG]} \BibitemShut {NoStop}%
\bibitem [{\citenamefont {Ye}\ \emph {et~al.}(2024)\citenamefont {Ye}, \citenamefont {Weng},\ and\ \citenamefont {Wu}}]{YE2024100003}%
  \BibitemOpen
  \bibfield  {author} {\bibinfo {author} {\bibfnamefont {C.-Y.}\ \bibnamefont {Ye}}, \bibinfo {author} {\bibfnamefont {H.-M.}\ \bibnamefont {Weng}},\ and\ \bibinfo {author} {\bibfnamefont {Q.-S.}\ \bibnamefont {Wu}},\ }\bibfield  {title} {\bibinfo {title} {Con-{CDVAE}: {A} method for the conditional generation of crystal structures},\ }\href {https://doi.org/https://doi.org/10.1016/j.commt.2024.100003} {\bibfield  {journal} {\bibinfo  {journal} {Comput. Mater. Today}\ }\textbf {\bibinfo {volume} {1}},\ \bibinfo {pages} {100003} (\bibinfo {year} {2024})}\BibitemShut {NoStop}%
\bibitem [{\citenamefont {Antunes}\ \emph {et~al.}(2024)\citenamefont {Antunes}, \citenamefont {Butler},\ and\ \citenamefont {Grau-Crespo}}]{Antunes2024}%
  \BibitemOpen
  \bibfield  {author} {\bibinfo {author} {\bibfnamefont {L.~M.}\ \bibnamefont {Antunes}}, \bibinfo {author} {\bibfnamefont {K.~T.}\ \bibnamefont {Butler}},\ and\ \bibinfo {author} {\bibfnamefont {R.}~\bibnamefont {Grau-Crespo}},\ }\bibfield  {title} {\bibinfo {title} {Crystal structure generation with autoregressive large language modeling},\ }\href {https://doi.org/10.1038/s41467-024-54639-7} {\bibfield  {journal} {\bibinfo  {journal} {Nat. Commun.}\ }\textbf {\bibinfo {volume} {15}},\ \bibinfo {pages} {10570} (\bibinfo {year} {2024})}\BibitemShut {NoStop}%
\bibitem [{\citenamefont {Merchant}\ \emph {et~al.}(2023)\citenamefont {Merchant}, \citenamefont {Batzner}, \citenamefont {Schoenholz}, \citenamefont {Aykol}, \citenamefont {Cheon},\ and\ \citenamefont {Cubuk}}]{gnome}%
  \BibitemOpen
  \bibfield  {author} {\bibinfo {author} {\bibfnamefont {A.}~\bibnamefont {Merchant}}, \bibinfo {author} {\bibfnamefont {S.}~\bibnamefont {Batzner}}, \bibinfo {author} {\bibfnamefont {S.~S.}\ \bibnamefont {Schoenholz}}, \bibinfo {author} {\bibfnamefont {M.}~\bibnamefont {Aykol}}, \bibinfo {author} {\bibfnamefont {G.}~\bibnamefont {Cheon}},\ and\ \bibinfo {author} {\bibfnamefont {E.~D.}\ \bibnamefont {Cubuk}},\ }\bibfield  {title} {\bibinfo {title} {Scaling deep learning for materials discovery},\ }\href {https://doi.org/https://doi.org/10.1038/s41586-023-06735-9} {\bibfield  {journal} {\bibinfo  {journal} {Nature}\ }\textbf {\bibinfo {volume} {624}},\ \bibinfo {pages} {80} (\bibinfo {year} {2023})}\BibitemShut {NoStop}%
\bibitem [{\citenamefont {Zeni}\ \emph {et~al.}(2025)\citenamefont {Zeni}, \citenamefont {Pinsler}, \citenamefont {Zügner}, \citenamefont {Fowler}, \citenamefont {Horton}, \citenamefont {Fu}, \citenamefont {Wang}, \citenamefont {Shysheya}, \citenamefont {Crabbé}, \citenamefont {Ueda}, \citenamefont {Sordillo}, \citenamefont {Sun}, \citenamefont {Smith}, \citenamefont {Nguyen}, \citenamefont {Schulz}, \citenamefont {Lewis}, \citenamefont {Huang}, \citenamefont {Lu}, \citenamefont {Zhou}, \citenamefont {Yang}, \citenamefont {Hao}, \citenamefont {Li}, \citenamefont {Yang}, \citenamefont {Li}, \citenamefont {Tomioka},\ and\ \citenamefont {Xie}}]{mattergen}%
  \BibitemOpen
  \bibfield  {author} {\bibinfo {author} {\bibfnamefont {C.}~\bibnamefont {Zeni}}, \bibinfo {author} {\bibfnamefont {R.}~\bibnamefont {Pinsler}}, \bibinfo {author} {\bibfnamefont {D.}~\bibnamefont {Zügner}}, \bibinfo {author} {\bibfnamefont {A.}~\bibnamefont {Fowler}}, \bibinfo {author} {\bibfnamefont {M.}~\bibnamefont {Horton}}, \bibinfo {author} {\bibfnamefont {X.}~\bibnamefont {Fu}}, \bibinfo {author} {\bibfnamefont {Z.}~\bibnamefont {Wang}}, \bibinfo {author} {\bibfnamefont {A.}~\bibnamefont {Shysheya}}, \bibinfo {author} {\bibfnamefont {J.}~\bibnamefont {Crabbé}}, \bibinfo {author} {\bibfnamefont {S.}~\bibnamefont {Ueda}}, \bibinfo {author} {\bibfnamefont {R.}~\bibnamefont {Sordillo}}, \bibinfo {author} {\bibfnamefont {L.}~\bibnamefont {Sun}}, \bibinfo {author} {\bibfnamefont {J.}~\bibnamefont {Smith}}, \bibinfo {author} {\bibfnamefont {B.}~\bibnamefont {Nguyen}}, \bibinfo {author} {\bibfnamefont {H.}~\bibnamefont {Schulz}}, \bibinfo {author} {\bibfnamefont {S.}~\bibnamefont {Lewis}}, \bibinfo {author}
  {\bibfnamefont {C.-W.}\ \bibnamefont {Huang}}, \bibinfo {author} {\bibfnamefont {Z.}~\bibnamefont {Lu}}, \bibinfo {author} {\bibfnamefont {Y.}~\bibnamefont {Zhou}}, \bibinfo {author} {\bibfnamefont {H.}~\bibnamefont {Yang}}, \bibinfo {author} {\bibfnamefont {H.}~\bibnamefont {Hao}}, \bibinfo {author} {\bibfnamefont {J.}~\bibnamefont {Li}}, \bibinfo {author} {\bibfnamefont {C.}~\bibnamefont {Yang}}, \bibinfo {author} {\bibfnamefont {W.}~\bibnamefont {Li}}, \bibinfo {author} {\bibfnamefont {R.}~\bibnamefont {Tomioka}},\ and\ \bibinfo {author} {\bibfnamefont {T.}~\bibnamefont {Xie}},\ }\bibfield  {title} {\bibinfo {title} {A generative model for inorganic materials design},\ }\href {https://doi.org/10.1038/s41586-025-08628-5} {\bibfield  {journal} {\bibinfo  {journal} {Nature}\ }\textbf {\bibinfo {volume} {639}},\ \bibinfo {pages} {624} (\bibinfo {year} {2025})}\BibitemShut {NoStop}%
\end{thebibliography}%

\end{document}